\documentclass[prd,10pt,aps,preprintnumbers,twocolumn,nofootinbib,superscriptaddress,showpacs,tightenlines,floatfix,amssymb,amsmath]{revtex4-1}
\usepackage{graphicx}
\usepackage{epstopdf}

\newcommand{\beqn}{\begin{eqnarray}}
\newcommand{\eeqn}{\end{eqnarray}}
\newcommand{\beqs}{\begin{subequations}}
\newcommand{\eeqs}{\end{subequations}}
\newcommand{\eq}[1]{(\ref{#1})}
\newcommand{\cD}{{\mathfrak D}}
\newcommand{\ext}{{\mathrm{ext}}}
\newcommand{\AS}{{\mathrm{AS}}}
\newcommand{\cL}{{\cal L}}
\newcommand{\cl}{{\ell}}

\newcommand{\cM}{{\cal M}}
\newcommand{\cA}{{\cal A}}
\newcommand{\dd}{\!{\mathrm d}}

\begin{document}

\title{Superconductivity of QCD vacuum in strong magnetic field}

\author{M. N. Chernodub}\thanks{On leave from ITEP, Moscow, Russia.}
\affiliation{CNRS, Laboratoire de Math\'ematiques et Physique Th\'eorique, \\
Universit\'e Fran\c{c}ois-Rabelais Tours, 
F\'ed\'eration Denis Poisson, Parc de Grandmont, 37200 Tours, France}
\affiliation{Department of Physics and Astronomy, University of
Gent, Krijgslaan 281, S9, B-9000 Gent, Belgium}

\begin{abstract}
We show that in a sufficiently strong magnetic field the
QCD vacuum may undergo a transition to a new phase where charged
$\rho^\pm$ mesons are condensed. In this phase the vacuum behaves as an
anisotropic inhomogeneous superconductor which supports superconductivity along
the axis of the magnetic field. In the directions transverse to the
magnetic field the superconductivity is absent. The
magnetic-field-induced anisotropic superconductivity -- which is
realized in the cold vacuum, i.e. at zero temperature and density -- is a consequence of a nonminimal coupling of
the $\rho$ mesons to the electromagnetic field. The onset of the
superconductivity of the charged $\rho^\pm$ mesons should also induce an
inhomogeneous superfluidity of the neutral $\rho^0$ mesons. We also argue that due to
simple kinematical reasons a strong enough magnetic field makes the
lifetime of the $\rho$ mesons longer by closing the main channels of
the strong decays of the $\rho$ mesons into charged pions.
\end{abstract}


\maketitle

\section{Introduction}

Properties of QCD matter subjected to very strong magnetic fields have recently attracted increasing interest of the community.
The interest is motivated by the possibility to create strong magnetic fields in the heavy-ion collisions
at RHIC and LHC. The strength of the magnetic field is estimated to be of the hadronic scale~\cite{Skokov:2009qp},
$eB \sim (1 \dots 15) \,m^2_\pi$, or even higher (here $m_\pi \approx 140\, {\mathrm{MeV}}$ is the pion mass). The duration of the
magnetic field ``flashes'' is expected to be rather short (a few fm$/c$).

Both analytical calculations~\cite{ref:B:effects,ref:splitting:1,ref:splitting:2} and lattice simulations~\cite{D'Elia:2010nq} indicate that
the QCD phase diagram is affected by the strong magnetic field. In particular, the external magnetic field
splits the chiral and deconfinement transitions~\cite{ref:splitting:1,ref:splitting:2}.  In a constant magnetic field of the
typical LHC magnitude, $eB \sim 15 m^2_\pi$~\cite{Skokov:2009qp}, the splitting between the critical temperatures of these transitions
reaches 10\,MeV~\cite{ref:splitting:1}.

In the quark-gluon plasma the strong magnetic field may also lead to the chiral magnetic effect~\cite{ref:CME}.
This effect generates an electric current of quarks along the magnetic field axis provided the densities of left-
and right-handed quarks are not equal.

In the cold matter the external magnetic field may create spatially inhomogeneous structures
which are made of quark condensates~\cite{ref:inhomogeneous:QCD}.

A recent lattice simulation has revealed that in the cold confinement phase the external magnetic field induces
nonzero electric conductivity along the direction of the field, thus transforming the QCD vacuum from an insulator
into an anisotropic conductor~\cite{Buividovich:2010tn}. In our paper we argue that there is a chance that
a stronger magnetic field may be able to make the QCD vacuum unstable towards creation of a superconducting state.
We would like to stress that we discuss here the electromagnetic superconductivity which should be distinguished
from the color superconductivity in the dense matter~\cite{Alford:2001dt}. We discuss a superconducting state
which may presumably be formed in the cold vacuum, i.e. at zero temperature and density.

Basically, we follow the works of Ambj\o rn, Nielsen and Olesen on two subjects: (i) on the condensate of color
magnetic flux tubes (``spaghetti states'')~\cite{Ambjorn:1979xi} created by an unstable gluonic mode in the QCD
vacuum~\cite{Nielsen:1978rm}; and (ii) on the condensation of the $W$-bosons in the standard electroweak model due to sufficiently
strong external magnetic field~\cite{Ambjorn:1988tm,Ambjorn:1988gb}. The key
idea Refs.~\cite{Ambjorn:1979xi,Nielsen:1978rm,Ambjorn:1988tm,Ambjorn:1988gb} is that the vacuum of charged
vector particles is unstable in the background of a sufficiently strong magnetic field provided these
particles have anomalously large gyromagnetic ratio $g=2$. The large value of $g$ guarantees that
the magnetic moment of such particles is too large to withstand a
spontaneous condensation at sufficiently strong external magnetic fields.

As we have mentioned, there are at least two examples of such instabilities.
A strong enough chromomagnetic field leads to the instability of the gluonic QCD vacuum since the gluon is the vector particle
with the (color) gyromagnetic ratio $g=2$~\cite{Nielsen:1978rm}. As a result of the instability, a spaghetti of the chromomagnetic flux
tubes is formed. These flux tubes tend to arrange themselves into a lattice structure
similar to the Abrikosov lattice which is realized in a mixed state of a type-II superconductor subjected to a near-critical
external magnetic field~\cite{Ambjorn:1979xi}.

The second example is suggested to be realized in
the standard electroweak model. The gyromagnetic ratio of the $W$ boson is also large, $g=2$,
so that in the strong magnetic field the vacuum of the electroweak theory is unstable towards formation
of the condensate of the $W$ bosons. The $W$ condensate is accompanied by a similar lattice vortex
state~\cite{Ambjorn:1988tm,Ambjorn:1988gb}. Note that in the
second example the external field is the electromagnetic field and not the color (gluon) one.

Our work is based on the fact that the $\rho$ meson is the charged vector particle with the gyromagnetic ratio $g=2$
so that this particle may condense in a background of strong enough magnetic field.
It important to stress that in all discussed cases of the spontaneous condensation -- we mentioned the gluons in QCD~\cite{Nielsen:1978rm},
the $W$ bosons in the electroweak theory~\cite{Ambjorn:1988gb,Ambjorn:1988tm}
and the $\rho$ mesons in QCD (this article) -- the condensation takes place in the {\it vacuum}
at zero temperature (as opposed to dense and/or hot environment).

The structure of the paper is as follows. In Section~\ref{sec:two} we outline the basic idea of the $\rho$-meson condensation.
In the same section we argue that the $\rho$ mesons are (at least, partially) stabilized by the strong magnetic field background.
This is an important property which should make the $\rho$ condensate ``intrinsically'' stable against decays of the $\rho$ mesons
(the $\rho$ mesons have a very short lifetime in the absence of the external fields). In Section~\ref{sec:model} we describe
the quantum electrodynamics of the $\rho$ mesons. Section~\ref{sec:GL} is devoted to a short overview of basic features
of the Ginzburg-Landau model of the superconductivity (homogeneity, isotropy, effects of the magnetic field,
the Abrikosov vortices, the Meissner effect, the London equations). In Section~\ref{sec:cond} we discuss the
same features in the superconducting state of condensed $\rho$ mesons in QCD and find a few similarities
and many surprising dissimilarities with the ordinary superconductivity. The last Section is devoted
to our conclusions.

\section{{\bf\large $\rho$} mesons in strong magnetic field: condensation and longer life}
\label{sec:two}

\subsection{Condensation of charged $\rho$ mesons}
\label{sec:condensation}

The basic idea of our work is as follows.
Consider a charged relativistic spin-$s$ particle moving in a background of an external magnetic field.
Without loss of generality we assume that the magnetic field $\vec B_\ext = (0,0,B_\ext)$ is directed along the $z$-axis,
$B_\ext \geqslant 0$ and we consider spatially uniform and time-independent external fields only.
The energy levels $\varepsilon$ of the free particle of the mass $m$ in the magnetic field are characterized by
three parameters: the nonnegative integer $n\geqslant 0$, the spin projection on the field's axis $s_z = -s, \dots, s$, and
the particle momentum along the field's axis, $p_z$:
\beqn
\varepsilon_{n,s_z}^2(p_z) = p_z^2+(2 n - 2 s_z + 1) eB_\ext + m^2\,.
\label{eq:energy:levels}
\eeqn
In this work we consider the charged particles, pions ($s=0$) and the vector particles,
$\rho$-mesons ($s=1$), for reasons that will be clear later. For a moment, we assume that these particles are free,
so that their (squared) minimal effective masses, corresponding to lowest energy states~\eq{eq:energy:levels}
with $p_z=0$, are, respectively:
\beqn
m_{\pi^\pm}^2(B_\ext) & = & m_{\pi^\pm}^2 + e B_\ext\,,
\label{eq:m2:pi:B} \\
m_{\rho^\pm}^2(B_\ext) & = & m_{\rho^\pm}^2 - e B_\ext\,.
\label{eq:m2:rho:B}
\eeqn
The zero-field vacuum masses of the $\pi^\pm$ and $\rho^\pm$ mesons are, respectively~\cite{ref:PDG},
\beqn
m_\pi = 139.6\,\mbox{MeV}\,, \qquad
m_\rho = 775.5\,\mbox{MeV}
\eeqn

\begin{figure}[!thb]
\begin{center}
\includegraphics[scale=0.48,clip=true]{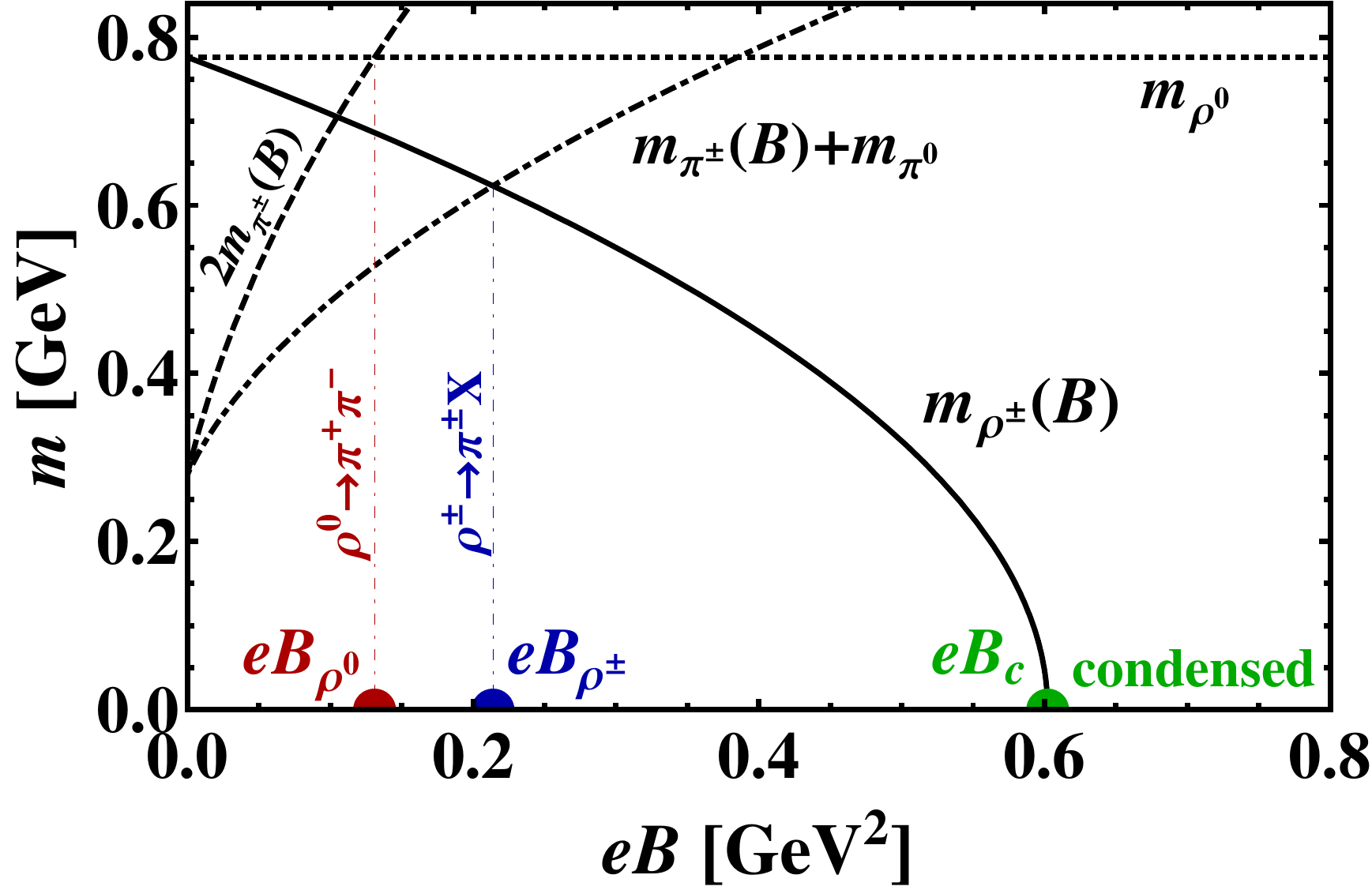}
\end{center}
\vskip -4mm
\caption{Masses of the lowest $\rho$-meson eigenstates and of the products of their dominant decay modes
as functions of the external magnetic field $B \equiv B_\ext$. The left red point and the middle blue point mark the
onsets of the ``$\pi^\pm$-stability regions'' for the neutral~\eq{eq:B:rho:0} and charged~\eq{eq:B:rho} $\rho$ mesons, respectively.
The right green point marks the critical field $B_c$ which corresponds to the onset of the $\rho^\pm$ condensation~\eq{eq:eBc}.}
\label{fig:critical}
\end{figure}

Equation~\eq{eq:m2:rho:B} implies that the lowest energy of the charged $\rho$-meson
in the external magnetic field may become purely imaginary if the magnetic field exceeds the following critical value
\beqn
B_c = \frac{m_\rho^2}{e} \approx 10^{16}\,\mbox{Tesla}\,,
\label{eq:eBc}
\eeqn
This observation indicates that the strong magnetic field ($B_\ext > B_c$) makes the QCD vacuum unstable towards
condensation of the charged $\rho$ mesons. This new QCD effect is very similar to the magnetic-field-induced
condensation of the $W$--bosons which was predicted by Ambj\o rn and Olesen~\cite{Ambjorn:1988tm,Ambjorn:1988gb}.
The behavior of the lowest mass~\eq{eq:m2:rho:B} of the charged $\rho^\pm$ meson in the region $0 \leqslant B_\ext \leqslant B_c$
is shown in Fig.~\ref{fig:critical} by the solid line.

The subtle point of Eq.~\eq{eq:m2:rho:B} [and of Eq.~\eq{eq:energy:levels} for $s=1$ as well] is that the gyromagnetic
ratio of the vector $\rho^\pm$ meson is set to be $g=2$. In fact, this $g$-factor is ``anomalously'' large
compared to the standard gyromagnetic ratio $g_{\mathrm{min}}=1$ of a charged vector particle which is minimally coupled to electromagnetism.
Notice, that it is the anomalous gyromagnetic ratio $g_W=2$ which drives the condensation of the $W$ bosons in the strong
magnetic field~\cite{Ambjorn:1988tm,Ambjorn:1988gb}. The large $g$-factor for $W$ boson is a direct consequence of the
non-Abelian nature of electroweak gauge group.

As for the $\rho$ mesons, the electrodynamics of these particles has also elements of a non--Abelian structure which is visible in
phenomenological Lagrangians~\cite{Weinberg:1968de,ref:QED:rho,ref:Effective:rho}. The vector dominance hypothesis~\cite{ref:Zumino}
as well as  the QCD sum rules~\cite{ref:Samsonov} point out that the $g$-factor of the $\rho$-mesons is 2. We discuss the quantum
electrodynamics for these vector mesons in more details in Sec.~\ref{sec:model}.

\subsection{Larger lifetime of charged and neutral $\rho$ mesons}

In the absence of the external magnetic field both the charged and neutral $\rho$ mesons are very unstable particles
characterized by the mean lifetime $\tau_\rho \approx 4.5 \times 10^{-24}\,{\mathrm{s}} \approx 1.35\,{\mathrm{fm}}/c$
which corresponds to the full width~\cite{ref:PDG}
\beqn
\Gamma_{\rho \to {\mathrm{all}}} = 149.1 \pm 0.8\,\mbox{MeV}\,.
\label{eq:Gamma:rho}
\eeqn
Thus, one may incorrectly conclude that if even the $\rho$--meson condensate is formed at
the strong magnetic fields, then it will anyway be unstable due to the intrinsic instability
of the $\rho$ mesons themselves. Below we show that this statement is incorrect.

\subsubsection{Charged vector mesons}

Consider first the charged vector mesons. All known decays of the $\rho^\pm$ mesons are going via the modes~\cite{ref:PDG}:
\beqn
\rho^\pm \to \pi^\pm X\,, \qquad X=\pi^0,\ \eta,\ \gamma,\ \pi\pi\pi\,.
\label{eq:rho:decays}
\eeqn
The fraction of the primary decay mode, $X=\pi^0$, is greater than $99\%$.

As the strength of the background magnetic field increases, the product of the decay, the charged pion
[which is always created in the known decay modes of the $\rho^\pm$ mesons~\eq{eq:rho:decays}]
becomes heavier~\eq{eq:m2:pi:B} while the decaying particle, the lowest state of the $\rho^\pm$ meson, becomes
lighter~\eq{eq:m2:rho:B}. Obviously, at a certain magnetic field $B_{\rho^\pm}$ the masses of the initial and final states in the
dominant channel, $\rho^\pm \to \pi^\pm \pi^0$, should become equal,
\beqn
m_{\rho^\pm}(B_{\rho^\pm}) = m_{\pi^\pm}(B_{\rho^\pm})+m_{\pi^0}\,,
\label{eq:equality:pm}
\eeqn
and the fast strong decays~\eq{eq:rho:decays} of the charged $\rho$
mesons should eventually become impossible due to obvious kinematical reasons. The strength of this ``$\pi^\pm$--stabilizing'' field is
approximately three times weaker\footnote{Here and below we always neglect the difference between the masses of
the charged $\pi^\pm$ and $\rho^\pm$ mesons, and their neutral counterparts, $\pi^0$
and $\rho^0$, respectively.} compared to the critical field of the $\rho$ condensation~\eq{eq:eBc},
\beqn
B_{\rho^\pm} = \frac{1}{2e} \bigl[m_\rho^2 - m^2_\pi - m_\pi (m^2_\pi + 2 m^2_\rho)^{\frac{1}{2}}\bigr] \simeq 0.36 B_c\,. \quad
\label{eq:B:rho}
\eeqn
The left and right hand sides of Eq.~\eq{eq:equality:pm} are shown by the solid and dot-dashed lines in
Fig.~\ref{fig:critical}. The point of the intersection of these lines gives us the critical field~\eq{eq:B:rho}.

At $B > B_{\rho^\pm}$ the charged $\rho$ mesons may in principle decay via other slower (and undetected so far) channels that avoid
fast gluon-mediated $\pi^\pm$ production. On the other side, the QCD string (which confines the quarks and antiquarks into mesons
and baryons) is partially stabilized by the external magnetic field~\cite{ref:stabilization}.
Thus, in the sufficiently strong magnetic field the allowed modes of the decays of the charged $\rho$ mesons should be much slower.
As the result, the lifetime of the $\rho^{\pm}$-mesons should be much longer compared to the lifetime of these particles in the
absence of the external magnetic field.

\begin{figure}[!thb]
\begin{center}
\includegraphics[scale=0.45,clip=true]{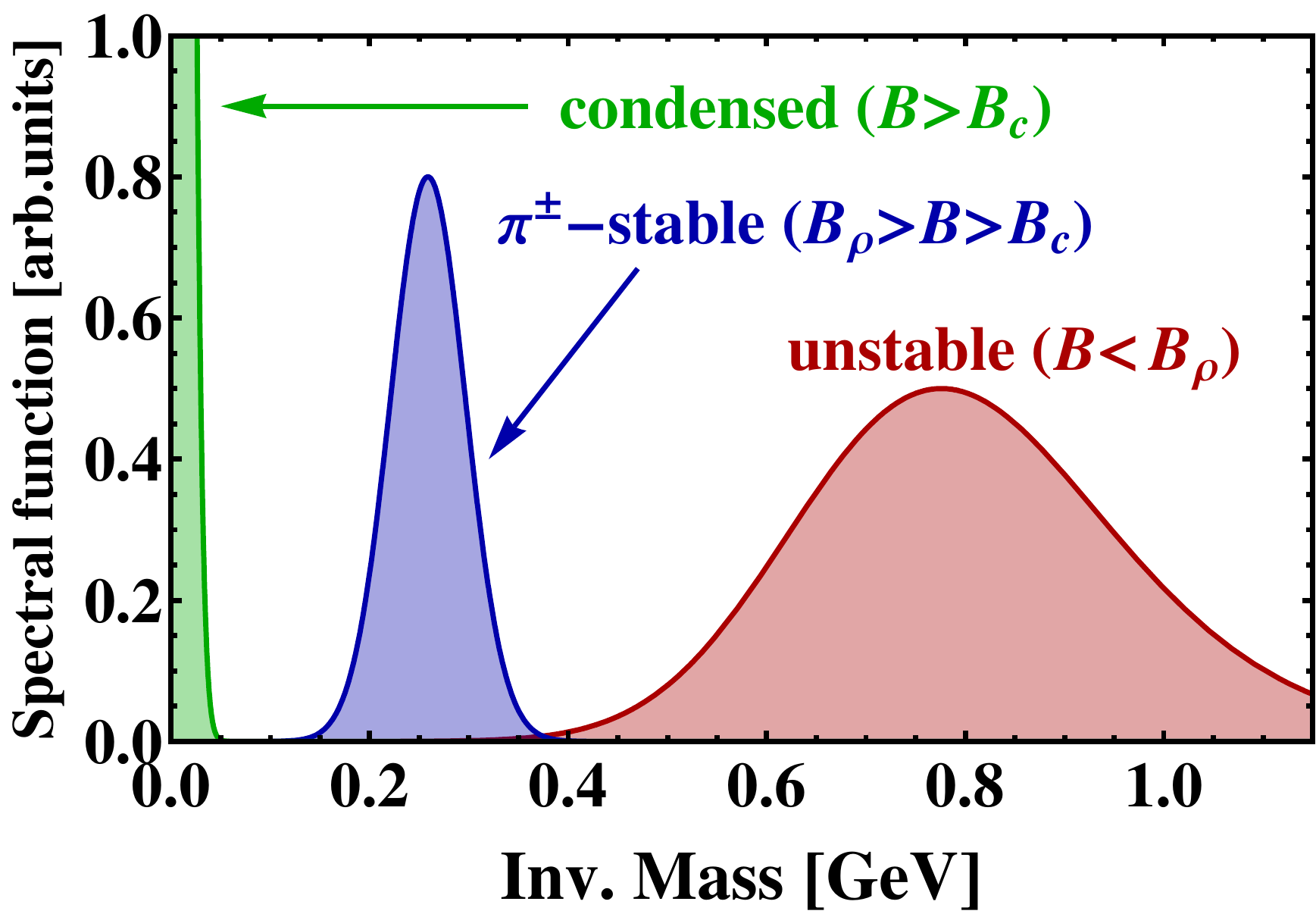}
\end{center}
\vskip -4mm
\caption{Prediction: a qualitative effect of the external magnetic field on the lowest-mass peak in the spectral function of
the $\rho^\pm$ mesons vs. the invariant mass. As the magnetic field strength $B \equiv B_\ext$ increases, the broad
peak in the unstable low-$B$ phase (right) turns into a much narrower peak in the $\pi^\pm$--stable phase (middle). At the onset of the
condensation of the $\rho$--mesons the peak transforms into the $\delta$-function--like singularity located at the vanishing
invariant mass. Features at higher invariant masses are not shown.}
\label{fig:spectral}
\end{figure}

One can also make a qualitative prediction for the behavior of spectral function of the charged
$\rho$ meson in the strong magnetic field. Expected behavior of the lowest-mass peak is
plotted in Fig.~\ref{fig:spectral} as a function of an invariant mass. At zero magnetic field the $\rho^\pm$ meson
is seen as a broad resonance (the right peak in Fig.~\ref{fig:spectral}). As we switch on the background magnetic field, the
single peak should split into multiple peaks corresponding to different levels of the charged vector particle ($s=1$)
in the external magnetic field~\eq{eq:energy:levels}.
The increase of the strength of the background magnetic field leads to the kinematical suppression of the $\rho$-meson
decay modes, and, consequently, to a narrower lowest--mass peak in the corresponding spectral function (the peak in the
middle of Fig.~\ref{fig:spectral}). At $B \geqslant B_c$, the onset of the condensation of the $\rho$--mesons
occurs. This effect can be seen as appearance of a singularity of the $\delta$-function--type located at the zero invariant mass
(the left peak in Fig.~\ref{fig:spectral}).

\subsubsection{Neutral vector mesons}

Similarly to its charged counterpart, the neutral $\rho^{0}$ meson should also be $\pi^\pm$--stabilized
in a sufficiently strong magnetic field background. The primary channel of the $\rho^0$ decay, $\rho^0 \to \pi^+ \pi^-$
(it corresponds to more than $99\%$ of the decays) becomes inoperative due to the same kinematical reasons provided
$B_\ext \geqslant B_{\rho^0}$, where
\beqn
m_{\rho^0}(B_{\rho^0}) = 2 m_{\pi^\pm}(B_{\rho^0})\,.
\label{eq:equality:0}
\eeqn
The $\rho^0$ mass is expected to be practically independent of the magnetic field\footnote{Here we
ignore a weak coupling of the magnetic field to the magnetic dipole moment of the $\rho^0$ meson.
This coupling makes the critical field~\eq{eq:B:rho:0} slightly stronger.},
$m_{\rho^0}(B) \simeq m_{\rho^0}(B=0)$, so that
\beqn
B_{\rho^0} = \frac{m_\rho^2 - 4 m^2_\pi}{4 e} \simeq 0.22 B_c\,.
\label{eq:B:rho:0}
\eeqn
The left and right hand sides of Eq.~\eq{eq:equality:0} are shown by the dashed and dotted lines in
Fig.~\ref{fig:critical}. The intersection of these lines occurs at the critical field~\eq{eq:B:rho:0}.

In the absence of the external magnetic field the neutral $\rho^0$ meson has also other decay channels which
do not involve the production of the charged $\pi^\pm$ pions. Such decay modes, however, are much slower compared to the primary
decay $\pi^\pm$-modes like $\rho^0 \to \pi^+ \pi^-$. For example, the most effective $\pi^\pm$-less decay of $\rho^0$
is $\rho^0 \to \pi^0 \gamma$, with the width
\beqn
\Gamma_{\rho^0 \to \pi^0 \gamma} = 0.089 \pm 0.012\,\mbox{MeV}\,,
\label{eq:Gamma:rho0}
\eeqn
which is more than three orders of magnitude narrower compared to the full width~\eq{eq:Gamma:rho}. In this paper we are not discussing
how the $\pi^\pm$-less decays are affected by the strong magnetic field. However, it is clear that the electromagnetically-driven
decay channels should be slower compared to the strongly-mediated ones. Thus, there are good kinematical reasons to believe
that the prolongation of the $\rho$-meson life -- induced by the strong magnetic field background -- should be substantial.

As for the evolution of the $\rho^0$ peak in the spectral function, we expect that the background magnetic field makes it narrower,
while its position is largely unaffected by the external field. As it will be clear from the results reported below, at $B_\ext > B_c$
we may expect an appearance of a singular peak at zero $\rho^0$-meson mass due to (quite weak, though) condensation
of the neutral $\rho^0$ mesons.

\subsubsection{Reversed decays and effects of chiral condensates}

The estimations of the values of the critical fields~\eq{eq:eBc}, \eq{eq:B:rho}, and \eq{eq:B:rho:0} are obviously approximate, as
one may expect systematic corrections coming from other effects of the strong magnetic field on the mass spectrum of the mesons.
For example, in our qualitative considerations we don't take into account effects of mixing of the $\rho^0$ meson
with the neutral $\omega$ and $\varphi$ mesons.
We also neglect influence of the magnetic field on the $\rho$ mesons and pions at the quark level.
However, the latter effect may be estimated, at least partially.
Indeed, the background magnetic field enhances the chiral symmetry breaking~\cite{ref:enhancement}.
According to a leading order of the chiral perturbation theory~\cite{ref:chiral:perturbation}
(confirmed by the results of the recent lattice simulations~\cite{Buividovich:2008wf})
the chiral condensate $\Sigma$ is a linearly increasing function of the strength of the external magnetic field $B_\ext$:
\beqn
\Sigma(B_\ext) = \Sigma(0) \Bigl(1 + \frac{\ln 2}{32 \pi^2 f_\pi^2} e B_\ext \Bigr)\,,
\label{eq:Sigma:B}
\eeqn
where $f_\pi = 92.4\,\mbox{MeV}$ is the pion decay constant. At the critical fields \eq{eq:B:rho:0}, \eq{eq:B:rho}
and \eq{eq:eBc} the corrections~\eq{eq:Sigma:B} to the chiral condensate are $3\%$, $6\%$, and $16\%$
respectively. We expect that uncertainties in our estimations of the critical values \eq{eq:B:rho:0}, \eq{eq:B:rho}
and \eq{eq:eBc} may be of the same scale at least.

One should also note that our considerations imply that at $B > B_{\rho^\pm}$ ($B > B_{\rho^0}$) the charged pions may decay into the
charged (neutral) $\rho$-mesons. The statement, that the presence of the strong enough magnetic field interchanges the decaying and
created particles, should not be disappointing. For example, it is known that the magnetic field may reverse the $\beta$-decay of the neutron
because at the background magnetic fields with the strength greater than  $5 \cdot 10^{14}\,{\mathrm{T}} \approx 0.1 \, m_e^2/e$ the proton becomes
heavier than the neutron. As a consequence of this effect, the proton may decay into the neutron by positron emission~\cite{Bander:1992ku}.

Summarizing this section, the charged and neutral $\rho$ mesons are very unstable particles provided the magnetic field is weaker
than the critical values $B_{\rho^\pm}$, Eq.~\eq{eq:B:rho}, and $B_{\rho^0}$, Eq.~\eq{eq:B:rho:0}, respectively (Fig.~\ref{fig:critical}).
We expect, however, that as the external field becomes stronger than these critical values, the corresponding $\rho$ mesons get stabilized
with respect to the vast majority of the strong decays which are going via the production of the $\pi^\pm$ mesons
(we call these regions of the magnetic field intensities as the ``$\pi^\pm$-stable'' phases both for the charged and neutral $\rho$ mesons).
If the background field surpasses the critical value~\eq{eq:eBc}, $B_\ext > B_c$, the condensation of the charged $\rho^\pm$ mesons should occur.
Below we show that at the same point $B_\ext = B_c$ the neutral $\rho^0$ mesons may simultaneously form an inhomogeneous superfluid.

\section{Electrodynamics of {\bf\large{$\rho$}} mesons}
\label{sec:model}

\subsection{The DSGS Lagrangian}

The self-consistent quantum electrodynamics for the $\rho$ mesons was recently constructed
by Djukanovic, Schindler, Gegelia and Scherer (DSGS) in Ref.~\cite{ref:QED:rho} starting from
an effective Lagrangian for vector mesons developed by Weinberg~\cite{Weinberg:1968de} long ago.
The chiral, Lorentz and discrete symmetries of the Weinberg Lagrangian
were extended to the Maxwellian $U(1)$ sector by adding all allowed interactions with electromagnetic fields.
In terms of the renormalized fields the bosonic part of the DSGS Lagrangian reads as follows~\cite{ref:QED:rho}:
\beqn
{\cal L} & = &-\frac{1}{4} \ F_{\mu\nu}F^{\mu\nu}
- \frac{1}{2} \ \rho^\dagger_{\mu\nu}\rho^{\mu\nu} + m_\rho^2 \ \rho_\mu^\dagger \rho^{\mu}
\label{eq:L:rho}\\
&& -\frac{1}{4} \ \rho^{(0)}_{\mu\nu} \rho^{(0) \mu\nu}+\frac{m_\rho^2}{2} \ \rho_\mu^{(0)}
\rho^{(0) \mu} +\frac{e}{2 g_s} \ F^{\mu\nu} \rho^{(0)}_{\mu\nu}\,,
\nonumber
\eeqn
where $A_\mu$ is the photon field,
$\rho_\mu = (\rho^{(1)}_\mu - i \rho^{(2)}_\mu)/\sqrt{2}$ and $\rho^{(0)}_\mu \equiv \rho^{(3)}_\mu$
are, respectively, the fields of the (negatively) charged and neutral vector mesons\footnote{We denote the
field of the neutral meson as $\rho^{(0)}(x)$ in order to discriminate it from the timelike component $\rho^0(x)$
of the charged $\rho^\pm$-meson field.} characterized by the mass~$m_\rho$.
The DSGS Lagrangian possesses the $U(1)$ gauge invariance:
\beqn
U(1)_{\mathrm{e.m.}}: \quad
\left\{
\begin{array}{lcl}
\rho^{(0)}_\mu(x) & \to & \rho^{(0)}_\mu(x)\,,\\
\rho_\mu(x) & \to & e^{i \omega(x)} \rho_\mu(x)\,,\\
A_\mu(x) & \to & A_\mu(x) + \partial_\mu \omega(x)\,. \quad
\end{array}
\right.
\label{eq:gauge:invariance}
\eeqn

The tensor quantities in \eq{eq:L:rho} are
\beqs
\beqn
F_{\mu\nu} & = & \partial_\mu A_\nu-\partial_\nu A_\mu\,,
\label{eq:F}\\
{f}^{(0)}_{\mu\nu} & = & \partial_\mu \rho^{(0)}_\nu-\partial_\nu \rho^{(0)}_\mu\,,
\label{eq:f0}\\
\rho^{(0)}_{\mu\nu}& = & {f}^{(0)}_{\mu\nu}
- i g_s(\rho^\dagger _\mu \rho_\nu-\rho_\mu\rho^\dagger _\nu)\,,
\label{eq:rho0}\\
\rho_{\mu\nu} & = & D_\mu \rho_\nu - D_\nu \rho_\mu\,,
\label{eq:rho}
\eeqn
\label{eq:rho:pm:coupling}
\eeqs
and the covariant derivative is
\beqn
D_\mu = \partial_\mu + i g_s \rho^{(0)}_\mu - ie A_\mu\,.
\eeqn

Equation~\eq{eq:rho:pm:coupling} indicates that $\rho^-_\mu \equiv \rho_\mu$ and $\rho^+_\mu \equiv \rho^\dagger_\mu$
meson fields carry the electric charges $- e$ and $+ e$ respectively (here $e = |e|$ is the elementary electric charge).
The coupling constant $g_s $ can be estimated~\cite{ref:QED:rho,ref:Effective:rho}
from the Kawarabayashi-Suzuki-Riadzuddin-Fayyazuddin relation~\cite{ref:KSRF}:
\beqn
g_s \equiv g_{\rho\pi\pi} = \frac{m_\rho}{\sqrt{2} f_\pi} = 5.88\,,
\label{eq:gs}
\eeqn
so that $g_s \gg e \equiv \sqrt{4 \pi \alpha_{\mathrm{e.m.}}} \approx 0.303$.

The most important fact for us is that the last term of the DSGS Lagrangian~\eq{eq:L:rho}
describes a nonminimal coupling of the $\rho$ mesons to the electromagnetic field.
This term has two parts,
\beqn
\delta \cL & = & \delta \cL^{(0)} + \delta \cL^{\mathrm{ch}}\,,  \\
\delta \cL^{(0)} & = & \frac{e}{2 g_s} \bigl(\partial_\mu \rho^{(0)}_\nu-\partial_\nu \rho^{(0)}_\mu\bigr) F^{\mu\nu} \,,
\label{eq:delta:L0}\\
\delta \cL^{\mathrm{ch}} & = & i e \rho_\mu\rho^\dagger_\nu \, F^{\mu\nu} \,.
\label{eq:delta:ch}
\eeqn
where the first part $\delta \cL^{(0)}$ corresponds to the coupling of the electromagnetic field to the magnetic dipole moment
of the $\rho^0$ meson, while the second part $\delta \cL^{\mathrm{ch}}$ describes the nonminimal coupling of the charged $\rho^\pm$-mesons
to the electromagnetic field. The presence of the former may lead to an instability of the vacuum of the neutral
vector particles ($\rho^{(0)}$ mesons in our case)~\cite{Bander:1991fg}, while the latter implies the anomalous
gyromagnetic ratio ($g = 2$) of the charged $\rho^\pm$ mesons, so that the magnetic dipole moment of the $\rho^\pm$
mesons is
\beqn
{\vec \mu}_{\rho^\pm} = \pm \, 2 \, \cdot \, \frac{e}{2 m_\rho} \, {\vec s}\,,
\label{eq:mu:rho}
\eeqn
(here ${\vec s}$ is the meson's spin). It is the coupling~\eq{eq:delta:ch} that plays a dominant effect in our
paper while the interaction~\eq{eq:delta:L0} makes a subleading contribution.

As we have already discussed in Section~\ref{sec:condensation},
spin-one particles with the gyromagnetic ratio $g = 2$ in strong enough external magnetic field should experience
a tachyonic instability towards development of a Bose-Einstein condensate. Since the condensed particles are charged, the
condensate should be superconducting, and this fact is our central observation which is discussed in details below.

\subsection{Equations of motion}
\label{sec:eqs:motion}

A variation of the DSGS Lagrangian~\eq{eq:L:rho} with respect to the electromagnetic potential $A_\mu$ provides
us with the Maxwell-type equation of motion,
\beqn
\partial^\nu F_{\nu\mu} = - J_\mu \,,
\label{eq:dF}
\eeqn
where the electric current $J_\mu$ contains two contributions,
\beqn
J_\mu = J^{\mathrm{ch}}_\mu + J^{(0)}_\mu\,,
\label{eq:Jmu}
\eeqn
coming from the charged and neutral mesons,
\beqs
\beqn
J^{\mathrm{ch}}_\mu & = & i e \bigl[\rho^{\nu\dagger} \rho_{\nu\mu} - \rho^\nu \rho^\dagger_{\nu\mu} +
\partial^\nu (\rho^\dagger_\nu \rho_\mu - \rho^\dagger_\mu \rho_\nu)\bigr]
\label{eq:Jch}\\
& \equiv & i e \bigl[(D_\mu \rho^\nu)^\dagger \rho_\nu - \rho^{\nu\dagger} D_\mu \rho_\nu\nonumber\\
& & + \partial^\nu (\rho^\dagger_\nu \rho_\mu - \rho^\dagger_\mu \rho_\nu)
+ \rho_\nu^\dagger D^\nu \rho_\mu - (D^\nu \rho_\mu)^\dagger \rho_\nu \bigr], \qquad \nonumber\\
J^{(0)}_\mu & = & - \frac{e}{g_s} \partial^\nu f^{(0)}_{\nu\mu}\,,
\label{eq:J0}
\eeqn
\label{eq:Jmu:2}
\eeqs
respectively. The currents \eq{eq:Jmu:2} are separately conserved:
\beqn
\partial^\mu J_\mu = \partial^\mu J^{\mathrm{ch}}_\mu = \partial^\mu J^{(0)}_\mu = 0\,.
\label{eq:dJ}
\eeqn

A variation of the DSGS Lagrangian~\eq{eq:L:rho} with respect to the field $\rho^{(0)}_\mu$ gives us the second equation of motion,
\beqn
\partial^\nu \rho^{(0)}_{\nu\mu} + m^2_\rho \rho^{(0)}_\mu {-} \frac{e}{g_s} \partial^\nu F_{\nu\mu}
{-} i g_s (\rho^\dagger_{\mu\nu} \rho^\nu - \rho_{\mu\nu} \rho^{\nu\dagger}) {=} 0\,. \qquad
\eeqn
It can be rewritten as follows [we used \eq{eq:dF}, \eq{eq:Jmu}, \eq{eq:Jmu:2}]:
\beqn
\bigl(\partial^\nu \partial_\nu + m^2_{\rho^{(0)}} \bigr) \rho^{(0)}_\mu - \partial_\mu \partial^\nu \rho^{(0)}_\nu
- \frac{g_s}{e} J^{\mathrm{ch}}_\mu = 0\,,
\label{eq:rho0:2}
\eeqn
so that Eq.~\eq{eq:dJ} gives us
\beqn
\partial^\mu \rho^{(0)}_\mu = 0\,.
\label{eq:drho0}
\eeqn

Equation~\eq{eq:rho0:2} provides us with the mass of the neutral $\rho^{(0)}$ meson:
\beqn
m_0 \equiv m_{\rho^{(0)}} = m_\rho \Bigl(1 - \frac{e^2}{g_s^2}\Bigr)^{-\frac{1}{2}}\,.
\label{eq:m:rho0}
\eeqn
Using Eqs.~\eq{eq:Jmu}, \eq{eq:Jmu:2} and \eq{eq:rho0:2} one can get
a well-known relation (emerged originally in the scope of vector dominance models long time ago~\cite{ref:Zumino})
between the electromagnetic current $J_\mu$ and the neutral meson field $\rho^{(0)}_\mu$:
\beqn
J_\mu = \frac{e m_0^2}{g_s} \rho_\mu^{(0)} \,,
\label{eq:rho0:J}
\eeqn
(notice that in our notations $e=|e|>0$).

The third equation of motion is
\beqn
D^\nu \rho_{\nu\mu} + m^2_\rho \rho_\mu + i (g_s \rho^{(0)}_{\mu\nu} - e F_{\mu\nu}) \rho^\nu = 0\,.
\label{eq:Drho}
\eeqn
Using the identity $[D_\mu, D_\nu] = i (g_s f^{(0)}_{\mu\nu} - e F_{\mu\nu})$, one gets
\beqn
\Bigl[\bigl(D^\alpha D_\alpha + m_\rho^2\bigr) g_{\mu\nu} - D_\mu D_\nu \qquad & &
\nonumber\\
+ i \bigl(g_s \rho^{(0)}_{\mu\nu} + g_s f^{(0)}_{\mu\nu} - 2 e F_{\mu\nu}\bigr)\Bigr] \rho^\nu & = & 0\,.
\label{eq:DDrho}
\eeqn
Equations~\eq{eq:Drho} and \eq{eq:rho0:J} imply that
\beqn
(\partial_\mu - i e A_\mu)\rho^\mu \equiv
\Bigl[D_\mu - \frac{i g^2_s}{e m^2_{\rho^{(0)}}} J_\mu\Bigr]\rho^\mu = 0\,.
\label{eq:Drho:0}
\eeqn

The linear part of Eq.~\eq{eq:DDrho} gives us the mass of the charged $\rho^\pm$ meson,
\beqn
m_{\rho^\pm} = m_\rho \,.
\label{eq:m:rho:ch}
\eeqn
The neutral vector $\rho^{(0)}$ meson is heavier compared to its charged counterpart $\rho^\pm$ .
According to Eqs.~\eq{eq:gs}, \eq{eq:m:rho0} and \eq{eq:m:rho:ch}, the difference in the masses is very small~\cite{ref:QED:rho},
\beqn
\delta m_{\rho} \equiv m_0 - m_{\rho^{\pm}} \simeq \frac{4 \pi \alpha_{\mathrm{e.m.}} f_\pi^2}{m_\rho} \approx 1 \, \mbox{MeV}\,.
\eeqn
This mass difference is consistent with the available experimental bounds~\cite{ref:PDG}.

\section{Example: Ginzburg--Landau model}
\label{sec:GL}

In the next Section~\ref{sec:cond} we analyze the condensation of the $\rho$ mesons in the strong magnetic field, starting from
the phenomenological field-theoretical DSGS Lagrangian~\eq{eq:L:rho}. However, before going into the details
of the $\rho$ condensation in QCD, it is very useful to discuss a few basic properties of
conventional superconductivity in the condensed matter physics. Below we concentrate on the Ginzburg-Landau (GL) model
which provides us with a simplest phenomenological description of the superconductivity.

\subsection{The Ginzburg--Landau Lagrangian}
The relativistic version of the GL Lagrangian for a superconductor is:
\beqn
\cL_{\mathrm{GL}} = - \frac{1}{4} F_{\mu\nu} F^{\mu\nu} + (\cD_\mu \Phi)^* \cD^\mu \Phi - \lambda (|\Phi|^2 - \eta^2)^2\,, \quad
\label{eq:L:GL}
\eeqn
where $\cD_\mu = \partial_\mu - i e A_\mu$ is the covariant derivative and
$\Phi$ is the complex scalar field carrying a unit\footnote{Without loss of generality,
it is convenient to consider the singly-charged bosons $\Phi$ instead of the usual doubly-charged bosons.} electric charge~$e$.

The ground state of the model~\eq{eq:L:GL} is characterized by the homogeneous condensate of the scalar field, $\Phi_0 = \langle \Phi\rangle$
with $|\Phi_0| = \eta$. In the condensed state the mass of the scalar excitation, $\delta \Phi = \Phi - \Phi_0$, and
the mass of the photon field $A_\mu$ are, respectively, as follows:
\beqn
m_\Phi = \sqrt{4 \lambda} \eta\,,
\qquad
m_A = \sqrt{2} e \eta\,.
\label{eq:masses}
\eeqn
The classical equations of motion of the GL model are
\beqn
\cD_\mu \cD^\mu \Phi + 2 \lambda (|\Phi|^2 - \eta^2) \Phi & = & 0\,,
\label{eq:GL:1}\\
\partial_\nu F^{\nu\mu} + J_{\mathrm{GL}}^\mu & = & 0\,,
\label{eq:GL:2}
\eeqn
where the electric current is
\beqn
J_{\mathrm{GL}}^\mu = - i e \bigl[\Phi^* \cD^\mu \Phi - (\cD^\mu \Phi)^* \Phi \bigr]\,.
\label{eq:GL:J}
\eeqn

\subsection{Destructive role of magnetic field}
\label{sec:GL:destructive}

The superconducting state in the GL model is completely destroyed ($\Phi = 0$) in a background of the
strong magnetic field $B_\ext$, if the strength of the field exceeds the critical value
\beqn
B^{\mathrm{GL}}_c = \frac{m^2_\Phi}{2e} \equiv \frac{2 \lambda}{e} \eta^2\,.
\label{eq:B:GL:c}
\eeqn

Let us assume that $B_\ext=F_{12}$ is the only nonvanishing component of the field strength tensor. Consider
the case when the uniform time-independent magnetic field $B_\ext$ is slightly smaller than the critical value~\eq{eq:B:GL:c},
$B<B^{\mathrm{GL}}_c$, so that
\beqn
1 - \frac{B_\ext}{B^{\mathrm{GL}}_c} \ll 1\,.
\label{eq:second}
\eeqn
Then the condensate is very small
\beqn
|\Phi_0(B)| \ll \eta\,,
\label{eq:first}
\eeqn
and Eq.~\eq{eq:GL:1} can be linearized:
\beqn
\bigr\{(\cD_1 - i \cD_2) (\cD_1 + i \cD_2) + e [B_c - B(x)]\bigl\} \Phi = 0\,,
\label{eq:D1D2}
\eeqn
where $B(x)$ is the field inside the superconductor (here we consider static and $z$-independent
solutions which correspond to a lowest energy of the system).
In the vicinity of the critical field $B \simeq B_\ext \simeq B_c$, so that Eq.~\eq{eq:D1D2}
reduces to the following equation for the condensate $\Phi$:
\beqn
\cD \Phi \simeq 0 \qquad \mbox{with} \quad \cD = \cD_1 + i \cD_2\,.
\label{eq:cD:phi}
\eeqn

The magnetic field {\it destroys} the superconductivity in the ordinary superconductor.
On the contrary, we show below that a strong enough magnetic field should {\it induce}
the superconductivity of the charged $\rho$ mesons in the QCD vacuum.

\subsection{Abrikosov lattice of vortices in mixed state}
\label{sec:GL:Abrikosov}

The GL model~\eq{eq:L:GL} admits a topological stringlike solution to the classical equations
of motion~\eq{eq:GL:1} and \eq{eq:GL:2}, which is known as the
Abrikosov vortex~\cite{Abrikosov:1956sx}.
The Abrikosov vortices are formed when the superconductors are subjected to external magnetic fields.

A single Abrikosov vortex carries the quantized magnetic flux (remember that we consider the condensed
bosons $\Phi$ which carry the electric charge $e$ and not $2e$):
\beqn
\int \dd^2 x_\perp \, B(x_\perp) = \frac{2 \pi}{e}\,,
\label{eq:quantized}
\eeqn
where the integral of the vortex magnetic field $B$ is taken over the
two-dimensional coordinates $x_\perp = (x_1,x_2)$ of the
plane which is transverse to the infinitely-long, strait and static vortex.
In the original solution, the scalar field of the unit-flux vortex
is singular at the vortex center,
\beqn
\Phi (x_\perp) \propto |x_\perp| e^{i \varphi} \equiv x_1 + i x_2\,,
\label{eq:Phi:vort}
\eeqn
where $\varphi$ is the azimuthal angle in the transverse plane,
and $|x_\perp|$ is the distance from the vortex center. Equation~\eq{eq:Phi:vort}
corresponds to small $|x_\perp|$: $m_\Phi |x_\perp| \ll 1$ and $m_A|x_\perp| \ll 1$.

In a type-II superconductor [in which $m_\Phi > m_A$, or,
according to \eq{eq:masses}, $2 \lambda > e^2$] the Abrikosov vortices repel each other. If the external field is strong enough
[but lower than the critical value \eq{eq:B:GL:c}] then multiple Abrikosov vortices are created.
Due to the mutual repulsion, the vortices arrange themselves in a regular structure known as
the Abrikosov lattice~\cite{ref:Landau,ref:Abrikosov}. Since the normal (nonsuperconducting) phase is
restored inside the vortices, the Abrikosov lattice corresponds to a ``mixed state'' of the superconductor,
in which both normal and superconducting states of matter are present.

There are various types of the Abrikosov lattices which are characterized by different energies~\cite{ref:Abrikosov}.
The stable lattice corresponds to a minimal energy of the system.
If the magnetic field $B_\ext$ approaches the critical magnetic field~\eq{eq:B:GL:c} from below, then
the simplest lattice type is given by the square lattice solution of Eq.~\eq{eq:cD:phi},
\beqn
\Phi(x_1,x_2) & = & C_0\, \exp\Biggl\{- \frac{(e B_\ext)^2}{2} x_1^2 \Biggr\}
\label{eq:Abrikosov:lattice}\\
& & \cdot \sum_{n = -\infty}^{+ \infty} \exp\Biggl\{- \pi n^2 + 2 \pi n \frac{x_1 + i x_2}{L_B}\Biggr\}\,.
\nonumber
\eeqn
In this equation the parameter $C_0$ is independent of the transverse coordinates $x_\perp$. The inter-vortex distance $L_B$
is expressed via the magnetic length $\cl_B$,
\beqn
L_B = \sqrt{2 \pi} \cl_B\,, \qquad \cl_B = \frac{1}{\sqrt{e B_\ext}}\,.
\label{eq:LB:GL}
\eeqn
The area of the elementary square cell (i.e. of a cell which contains one Abrikosov vortex)
is $L^2_B \equiv 2 \pi \cl_B$. The absolute
value of the condensate, $|\Phi(x_\perp)|$, has a square symmetry in the solution~\eq{eq:Abrikosov:lattice} and
the vortices are located at the sites of the square lattice,
\beqn
\frac{x_i}{L_B} = n_i + \frac{1}{2}\,, \qquad n_i \in {\mathbb{Z}}\,, \quad i=1,2\,.
\label{eq:vort:locations}
\eeqn
In this case the distance between the vortex centers is~$L_B$. At the points~\eq{eq:vort:locations}
the condensate $\Phi(x_1,x_2)$ vanishes exactly and in the vicinity of these points the scalar
field~\eq{eq:Abrikosov:lattice} follows the behavior of Eq.~\eq{eq:Phi:vort}.

As we will see below, the pure superconducting state cannot be formed
in the $\rho$ meson superconductor contrary to the ordinary superconductor.
Instead, the Abrikosov lattice state is created.

\subsection{Homogeneous isotropic superconductivity}
\label{sec:GL:homogeneous}

Let us now apply a very weak external electromagnetic field to the superconductor.
Neglecting effect of the external field on the condensate $\Phi_0$, one gets from~\eq{eq:GL:J}:
\beqn
\partial^{\mu} J_{\mathrm{GL}}^{\nu} - \partial^{\nu} J_{\mathrm{GL}}^{\mu} = - m^2_A F^{\mu\nu}\,,
\label{eq:dJ:London}
\eeqn
where $m_A$ is given in Eq.~\eq{eq:masses}. Setting $\mu{=}0$ and $\nu{=}i$ in Eq.~\eq{eq:dJ:London}
one gets the first London relation for a locally neutral [$J_0(x) = 0$] superconductor
\beqn
\frac{\partial {\vec J}_{\mathrm{GL}}}{\partial t} = m_A^2 {\vec E}\,,
\label{eq:London:GL}
\eeqn
where $E^i \equiv - F^{0i}$ is the time-independent and uniform electric field. Equation~\eq{eq:London:GL}
implies a linear growth of the electric current in external electric field, thus indicating a vanishing
electric resistance of the superconducting state.

In the long-wavelength limit, $|\vec q | \to 0$, the weak electric
field ${\vec E}(\vec x, t) = {\vec E}_0 e^{i (\vec x \cdot \vec q - \omega t)}$
induces the local current
\beqn
J_k(\vec x, t;\omega) = \sum_{k=1}^3 \sigma_{kl}(\omega) E_l(\vec x, t)\,,
\label{eq:J:London}
\eeqn
where $\sigma_{kl} = {\mathrm{Re}}\, \sigma_{kl} + i \, {\mathrm{Im}}\, \sigma_{kl}$ is the complex electric conductivity.
The London equation~\eq{eq:London:GL} indicates that
\beqn
\sigma_{kl}(\omega) = \sigma^{\mathrm{sing}}_{kl}(\omega) + \sigma^{\mathrm{reg}}_{kl}(\omega)\,,
\label{eq:sigma:London:1}
\eeqn
where the first contribution is a singular isotropic part associated with the superconducting state:
\beqn
\sigma^{\mathrm{sing}}_{kl} (\omega) = \frac{\pi m^2_A}{2} \Bigl[\delta(\omega) + \frac{2 i}{\pi \omega}\Bigr] \delta_{kl}\,.
\label{eq:sigma:London:2}
\eeqn
The regular part $\sigma^{\mathrm{reg}}$ accounts for all other (nonsuperconducting) contributions to the conductivity.

It is clear that the superconductivity described by Eq.~\eq{eq:London:GL} is homogeneous (it is independent of the spatial coordinate)
and isotropic (it is independent of the direction). On the contrary, we will see below that a strong enough magnetic field
induces {\it inhomogeneous} and {\it anisotropic} superconductivity of the charged $\rho$ mesons in the QCD vacuum.

\subsection{Meissner effect}
\label{sec:GL:Meissner}

The spatial components of Eq.~\eq{eq:dJ:London} give us the second London relation:
\beqn
{\vec \partial} \times {\vec J}_{\mathrm{GL}} = - m^2_A {\vec B}\,,
\label{eq:London:GL:2}
\eeqn
so that in the absence of the external electric field ($\vec E = 0$)
one of the Maxwell equations~\eq{eq:GL:2}, ${\vec J}_{\mathrm{GL}} = {\vec \partial} \times {\vec B}$,
implies
\beqn
(- \Delta + m^2_A) {\vec B} = 0\,.
\label{eq:Meissner}
\eeqn
This equation indicates that the photon inside the superconductor acquires the mass $m_A$, Eq.~\eq{eq:masses}.
Consequently, the superconductor tends to expel the external magnetic field (``the Meissner effect'').
Physically, the Meissner effect is realized because the external magnetic field induces the
circulating superconducting currents~\eq{eq:London:GL:2} inside the superconductor.
These currents, in turn, screen the external magnetic field since they induce their own magnetic field
which is opposite to the external one (here we always assume that $B_\ext < B_c$).

A weak magnetic field which is parallel to the boundary of the superconductor is always screened inside the bulk
of the superconductor. The perpendicular magnetic field may penetrate the superconductor and create a mixed phase
of the Abrikosov vortices.

As we will see below, the second London equation \eq{eq:London:GL:2}
is not realized in the superconducting phase of the QCD vacuum contrary
to the conventional superconductor. Consequently, the Meissner effect cannot
be formulated in a selfconsistent way in the suggested superconducting phase of QCD.

\section{Condensation of {\bf\large{$\rho$}} mesons}
\label{sec:cond}

\subsection{Homogeneous approximation}
\label{sec:homogeneous}

The energy density of the DSGS model~\eq{eq:L:rho} is
\beqn
& & \hskip -4mm \epsilon \equiv T_{00} = \frac{1}{2} F_{0i}^2 + \frac{1}{4} F_{ij}^2
+ \frac{1}{2} \bigl(\rho_{0i}^{(0)}\bigr)^2 + \frac{1}{4} \bigl(\rho_{ij}^{(0)}\bigr)^2
\nonumber\\
& &  + \frac{m_\rho^2}{2}\Bigl[\bigl(\rho_{0}^{(0)}\bigr)^2 + \bigl(\rho_{i}^{(0)}\bigr)^2\Bigr]
+ \rho_{0i}^\dagger \rho_{0i} + \frac{1}{2} \rho_{ij}^\dagger \rho_{ij}
\quad \label{eq:energy:density}\\
& &
+ m_\rho^2\bigl(\rho_{0}^\dagger \rho_{0} + \rho_{i}^\dagger \rho_{i}\bigr)
- \frac{e}{g_s} F_{0i} \rho^{(0)}_{0i} - \frac{e}{2 g_s} F_{ij} \rho^{(0)}_{ij}\,,
\nonumber
\eeqn
were $T_{\mu\nu}$ is the energy-momentum tensor,
\beqn
T_{\mu\nu} = 2 \frac{\partial {\cal L}}{\partial g^{\mu\nu}} - {\cal L}\, g_{\mu\nu}\,.
\eeqn

In order to understand the phase structure of the $\rho$ mesons in the background magnetic field,
it is useful to study first the homogeneous field approximation. To this end we
ignore the kinetic terms $\partial_\mu \rho_\nu^{(0)} = 0$ and $D_\mu \rho_\nu = 0$
in Eq.~\eq{eq:energy:density}. The remaining (potential) part of the energy density
in the external uniform magnetic field $B_\ext$ is:
\beqn
& & \epsilon_0\bigl(\rho_\mu,\rho^{(0)}_\nu\bigr) = \frac{1}{2} B^2_\ext + \frac{g_s^2}{4}
\bigl[i \bigl(\rho_\mu^\dagger \rho_\nu - \rho_\nu^\dagger \rho_\mu\bigr)\bigr]^2
\label{eq:epsilon0}\\
& & \hskip 7mm
+ i e B_\ext \bigl(\rho_1^\dagger \rho_2 - \rho_2^\dagger \rho_1\bigr)
+ \frac{m_\rho^2}{2} {\bigl(\rho_\mu^{(0)}\bigr)}^2
+ m_\rho^2 \rho_\mu^\dagger \rho_\mu\,. \quad
\nonumber
\eeqn
where the sums over silent indices are written in the Euclidean metric,
$O_\mu^2 \equiv \sum_{\mu=0}^3 O_\mu O_\mu$. We remind that
we always take $B_\ext \equiv F_{12}>0$, and in this Section $F_{0 i} = F_{3 i} = 0$.

The ground state of the model can be found via the minimization of the potential energy~\eq{eq:epsilon0}
with respect to the meson fields. To this end we notice that the field of the neutral meson is vanishing
at the energy minimum, $\rho_\mu^{(0)}=0$. Then, the quadratic part of Eq.~\eq{eq:epsilon0} becomes as follows:
\beqn
\epsilon_0^{(2)}(\rho_\mu) & = & i e B_\ext \bigl(\rho_1^\dagger \rho_2 - \rho_2^\dagger \rho_1\bigr) + m_\rho^2 \rho_\mu^\dagger \rho_\mu
\nonumber\\
& = & \sum_{a,b=1}^2\rho_a^\dagger \cM_{ab} \rho_b + m_\rho^2 (\rho_0^\dagger \rho_0 + \rho_3^\dagger \rho_3)\,.
\label{eq:rho:mu}
\eeqn
The Lorentz components $\rho_1$ and $\rho_2$
possess the non--diagonal mass matrix
\beqn
\cM =
\left(
\begin{array}{cc}
m_\rho^2 & i e B_\ext \\
- i e B_\ext & m_\rho^2
\end{array}
\right)\,.
\label{eq:cM}
\eeqn
The eigenvalues $\mu_{\pm}$ and the corresponding eigenvectors $\rho_{\pm}$
of the mass matrix \eq{eq:cM} are, respectively, as follows:
\beqn
\mu_{\pm}^2 = m_\rho^2 \pm e B_\ext\,,
\qquad
\rho_{\pm} = \frac{1}{\sqrt{2}} (\rho_1 + i \rho_2)\,.
\label{eq:rho:diag}
\eeqn
The mass terms for $\rho_0$ and $\rho_3$ components are diagonal in \eq{eq:rho:mu} and their
prefactors $m_\rho^2$ are unaltered by the external magnetic field.

It is clear from Eq.~\eq{eq:rho:mu} that in terms of the ``longitudinal'' components $\rho_0$ and $\rho_3$,
the ground state of the model corresponds to $\rho_0 = \rho_3 = 0$ at any value of the magnetic field.
We express the transverse components $\rho_{1,2}$ via the eigenvalues and eigenvectors \eq{eq:rho:diag}
of the mass matrix~\eq{eq:cM}, and then we get for (the potential part of) the energy density~\eq{eq:energy:density}
the following expression:
\beqn
\epsilon_0(\rho_+,\rho_-) = \frac{1}{2} B^2_\ext & {+} & \frac{g^2_s}{2} \bigl(|\rho_+|^2 - |\rho_-|^2\bigr)^2
\nonumber\\
& & + \mu_+^2 |\rho_+|^2 + \mu_-^2 |\rho_-|^2\,.
\label{eq:epsilon:0:1}
\eeqn
Since $\mu_+^2> 0$ regardless of the value of the magnetic field $B_\ext$, the ground state corresponds to
$\rho_+ = 0$. In turn, this means that $\rho_- \equiv \sqrt{2} \rho$ and
\beqn
\rho_1 = - i \rho_2 = \rho\,, \qquad \rho_0 = \rho_3 = 0\,,
\label{eq:asymm:ground:state}
\eeqn
where $\rho$ is a scalar complex field. In terms of the new field $\rho$
the energy density~\eq{eq:epsilon:0:1} takes the simple form:
\beqn
\epsilon_0 (\rho) = \frac{1}{2} B^2_\ext + 2 (m_\rho^2 - e B_\ext) \, |\rho|^2 + 2 g_s^2 \, |\rho|^4\,.
\label{eq:epsilon:0:2}
\eeqn
Thus, we get the familiar Mexican-hat potential which describes various spontaneously broken systems. In particular,
the same potential appears in the GL model of superconductivity~\eq{eq:L:GL}.

\begin{figure}[!thb]
\begin{center}
   \includegraphics[scale=0.42,clip=true]{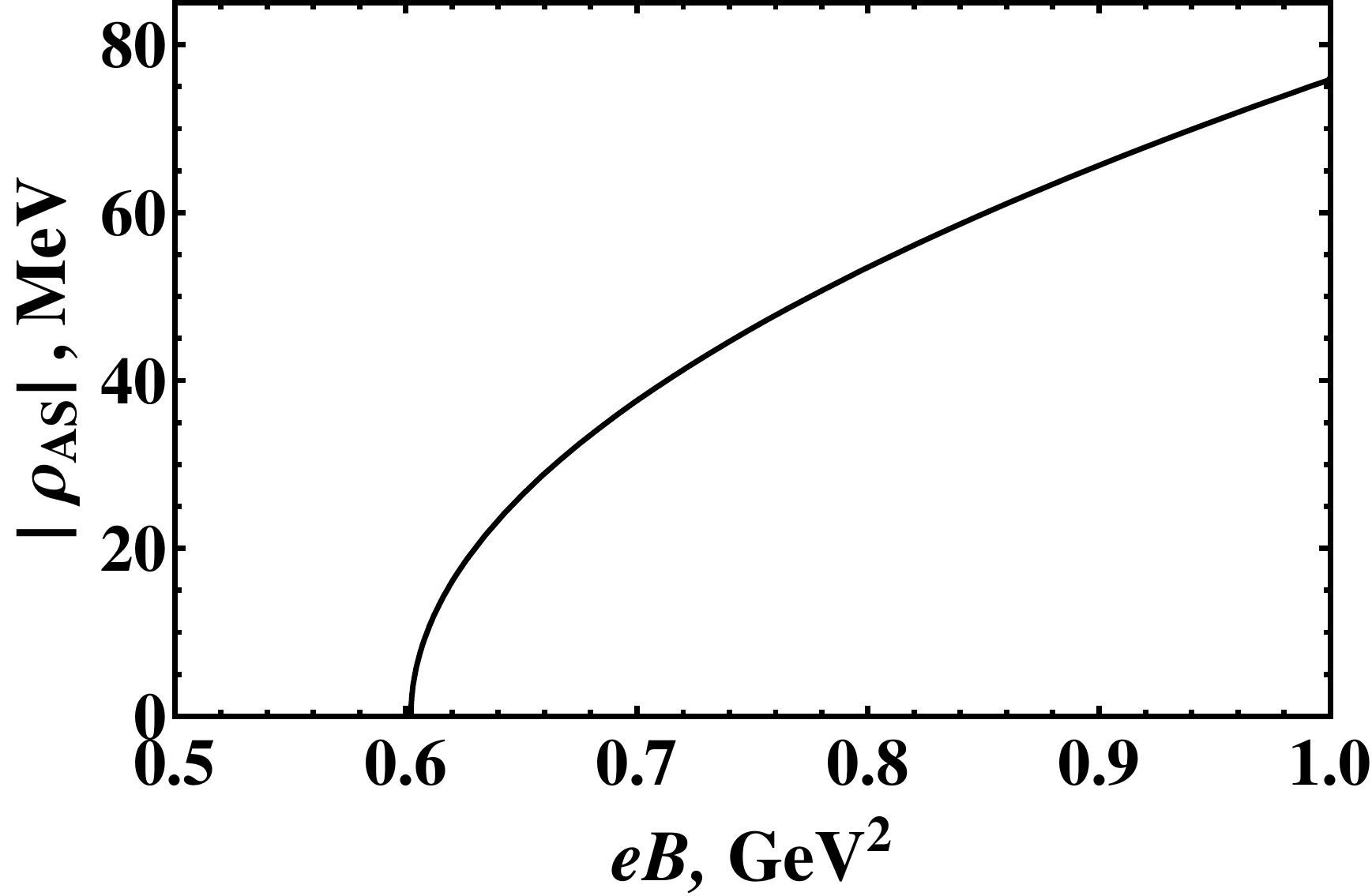}
\end{center}
\vskip -4mm
\caption{The condensate $|\rho_{AS}|$ of the charged $\rho^\pm$
mesons as a function of the external magnetic field $B \equiv B_\ext$ at the ground state.
This single curve describes both the uniform condensate $|\rho_{AS}| \equiv |\rho|_{{}_0}$
in the homogeneous approximation~\eq{eq:norm:rho2:h} and the mean-cell value
$|\rho_{AS}| \equiv |\rho_{AS}|_{{}_\cA}$ of the inhomogeneous condensate~\eq{eq:norm:rho2} in the weak-amplitude approximation.}
\label{fig:condensate}
\end{figure}

The ground state of the model~\eq{eq:epsilon:0:2} depends on the value of the external magnetic field:
if the field strength is weaker than the critical value $B_c = m^2_\rho/e$, Eq.~\eq{eq:eBc}, then the potential is trivial,
while if $B_\ext>B_c$ then we get a nontrivial ground state:
\beqn
|\rho|_{{}_0} =
\left\{
\begin{array}{lcl}
\sqrt{\frac{e(B_\ext - B_c)}{2 g^2_s}}\,, & \quad & B_\ext  \geqslant B_c\,,\\
0\,, & \quad & B_\ext < B_c\,,
\end{array}
\right.
\label{eq:norm:rho2:h}
\eeqn
(the subscript ``0'' in $|\rho|_{{}_0}$ indicates that we consider the homogeneous-field approximation).
In Fig.~\ref{fig:condensate} we plot the behavior of the condensate~\eq{eq:norm:rho2:h} as the function of
the external magnetic field $B_\ext$. The value of the condensate follows
a typical behavior of an order parameter for a second order phase transition at $B_\ext = B_c$.

In Fig.~\ref{fig:condensate} the subscript ``AS'' in $|\rho_{AS}|$ stands for the ``anisotropic superconductor''.
Indeed, the scalar field $\rho(x)$ enjoys the gauge symmetry~\eq{eq:gauge:invariance} of its vector predecessor~$\rho_\mu(x)$,
\beqn
U(1)_{\mathrm{e.m.}}: \qquad \rho(x) & \to & e^{i \omega(x)} \rho(x)\,.
\label{eq:gauge:invariance:rho}
\eeqn
The formation of the nontrivial ground state $\rho$ in the strong external magnetic field $B_\ext  \geqslant B_c$
breaks spontaneously the gauge symmetry \eq{eq:gauge:invariance} and forms, consequently, a superconducting state.
The superconductor should exhibit spatially anisotropic properties due to spatially anisotropic condensate~\eq{eq:asymm:ground:state}.
This issue will be discussed in details later.

Note that in the presence of the background magnetic field ${\vec B}_\ext$ the rotational group $SO(3)_{\mathrm{rot}}$ is explicitly broken
to its $O(2)_{\mathrm{rot}}$ subgroup generated by rotations around the axis of the magnetic field. In the homogeneous approximation,
the ground state~\eq{eq:asymm:ground:state} is transformed under the global $O(2)_{\mathrm{rot}}$ rotations as follows:
\beqn
O(2)_{\mathrm{rot}}: \qquad \rho(x) & \to & e^{i \varphi} \rho(x)\,,
\label{eq:rot:invariance:rho}
\eeqn
where $\varphi$ is the azimuthal angle of the rotation in the transverse plane. Thus, the ground state~\eq{eq:asymm:ground:state}
is invariant under a combination of the global transformation from the gauge group~\eq{eq:gauge:invariance:rho} and the
global rotation around the field axis~\eq{eq:rot:invariance:rho}
provided the parameters of these transformations are related (``locked'') to each other as follows: $\omega(x) = - \varphi$. In
analogy with the color superconductivity~\cite{Alford:2001dt} one can say that the ground state ``locks'' the residual
rotational symmetry with the electromagnetic gauge symmetry:
\beqn
U(1)_{\mathrm{e.m.}} \times O(2)_{\mathrm{rot}} \to U(1)_{\mathrm{locked}}\,.
\label{eq:locking}
\eeqn
Below we will see that the inhomogeneities of the condensate break the locked group~\eq{eq:locking} further to the group
of discrete rotations of the vortex lattice.

In the ground state~\eq{eq:norm:rho2:h} the potential energy~\eq{eq:rho:mu} has the form:
\beqn
\frac{\varepsilon_0 (|\rho|=|\rho|_{{}_0})}{\varepsilon_0 (|\rho|=0)} =
\left\{\!\!
\begin{array}{lcl}
1 - \frac{e^2}{g_s^2} \Bigl(1 - \frac{B_c}{B_\ext}\Bigr)^2, & & B_\ext  \geqslant B_c,\\
1, & & B_\ext < B_c,
\end{array}
\right.\qquad
\label{eq:epsilon:h}
\eeqn
Obviously, for a strong magnetic field $B_\ext \geqslant B_c$, the condensed state
has lower energy compared to the energy $\varepsilon_0 (|\rho|=0) = B_\ext^2/2$ of the
normal (noncondensed) state.

Thus, we observed that the condensation of the $\rho^\pm$ mesons in the QCD vacuum should be very different from
the condensation of the Cooper pairs $\Phi$ in the standard superconductor which is described by the phenomenological GL model~\eq{eq:L:GL}.
Indeed, in Section~\ref{sec:GL:destructive}
we have illustrated the destructive role of the strong magnetic field on the conventional superconductivity.
On the contrary, in this Section we have found that the strong enough magnetic field enforces the $\rho$-meson superconductivity.

\subsection{Two-dimensional equations of motion}
\label{sec:2d}

In order to study the properties of the emerged superconductor in more details we
should definitely go beyond the homogeneous approximation.
The inhomogeneous state can be treated with the full system of the $3+1$ dimensional equations of motion for the $\rho$-meson
fields which were discussed in Section~\ref{sec:eqs:motion}.
We notice, however, that a wavefunction of the lowest energy state of a free particle in a uniform static
magnetic field is independent on the coordinate $x_3$ which is longitudinal to the magnetic field. The
dependence on the time coordinate $x_0$ comes as a trivial phase factor only. The Abrikosov lattice solution
in the type-II superconductors is also
known to be independent of $x_0$ and $x_3$ coordinates (Section~\ref{sec:GL:Abrikosov}). These well-known properties suggest
us to concentrate on $x_0$- and $x_3$-independent solutions to the classical equation of motions for the $\rho$ mesons.
To this end we choose the complex coordinate $z = x_1 + i x_2$ where $x_\perp = (x_1,x_2)$ are the coordinates
in the spatial plane which is transverse of the magnetic field axis. We define
the complex variables
\beqn
{\cal O} = {\cal O}_1 + i {\cal O}_2\,, \qquad
{\overline{\cal O}} = {\cal O}_1 - i {\cal O}_2
\eeqn
for the fields ${\cal O} = J^{(0)}$, $J$, $\rho^{(0)}$, $A$, and for the derivative ${\cal O} = \partial$.
It is also convenient to introduce two covariant derivatives:
\beqn
D \equiv D_1 + i D_2 = \cD + i g_s \rho^{(0)}\,, \qquad \cD = \partial - i e A. \qquad
\label{eq:covariant}
\eeqn
For the sake of convenience we use below both $x_\perp$ and $z$ notations interchangeably, so that the two-dimensional Laplacian, for example,
can be written in the three different ways: $\partial \bar \partial \equiv \partial_\perp^2 \equiv \partial_1^2 + \partial_2^2$.

Our homogeneous field analysis (Section~\ref{sec:homogeneous})
suggests that the charged currents should be chosen in the form
\beqn
\rho_0=\rho_3 = 0\,,
\qquad
\rho_1= - i \rho_2 = \rho(z)\,,
\label{eq:rho:ansatz}
\eeqn
where $\rho$ is a complex field\footnote{In a strong field limit one can show that
due to presence of inhomogeneities
the ansatz \eq{eq:rho:ansatz} may be generalized :
$\rho_1 = \rho(z)+\xi(z)$, $\rho_2 = i [\rho(z) - \xi(z)]$.
In our analysis we ignore the subleading field $\xi$ because its amplitude is suppressed by the factor $e/g_s \ll 1$.}.

The magnetic field~\eq{eq:F} and the field-strength of the neutral vector bosons~\eq{eq:f0} are as follows
\beqn
F_{12} \equiv {\mathrm{Im}}(\bar{\partial} A) & = & B(z)\,,
\label{eq:B:2d}
\\
f^{(0)}_{12} \equiv {\mathrm{Im}}(\bar{\partial} \rho^{(0)}) & = & C(z)\,. \qquad
\label{eq:C:2d}
\eeqn
Notice, that despite the external magnetic field $B_\ext$ is assumed to be uniform,
the magnetic field~\eq{eq:B:2d} of the classical solution may be (and, in fact, will be) inhomogeneous.
The tensor quantities~\eq{eq:rho0} and \eq{eq:rho} take, respectively, the following form (we omit the argument $z$ hereafter):
\beqn
\rho^{(0)}_{12} = C + 2 g_s |\rho|^2\,, \qquad
\rho_{12} = i D \rho\,.
\eeqn

The charged and neutral components of the current~\eq{eq:Jmu:2} become simple expressions, respectively:
\beqn
J^{\mathrm{ch}} = 2 i e \bigl(\rho^\dagger D \rho + \partial |\rho|^2 \bigr)\,,
\qquad
J^{(0)} & = & i \frac{e}{g_s}\partial C\,.
\eeqn
The conservation law for the charged current~\eq{eq:dJ},
${\mathrm{Im}} \bigl\{\bar{\partial} \bigl[\rho^\dagger \cD \rho - \rho (\bar{\cD} \rho)^\dagger \bigr] \bigr\} = 0$,
is satisfied automatically due to relation~\eq{eq:Drho:0},
\beqn
\cD \rho = 0\,,
\label{eq:cDrho0}
\eeqn
[we also used the identity $\partial |\rho|^2 \equiv \rho^\dagger D \rho + ({\bar{D}} \rho)^\dagger \rho$].

Equations \eq{eq:dF}, \eq{eq:rho0:2}, \eq{eq:Drho} reduce, respectively, to
\beqn
g_s \partial B + i e m^2_0 \rho^{(0)} & = & 0,
\label{eq:one}\\
\bigl( - \bar{\partial}\partial + m^2_0 + 2 g_s^2 |\rho|^2\bigr) \rho^{(0)} - 2 i g_s \partial |\rho|^2 & = & 0, \qquad
\label{eq:two}\\
\bigl[- \bar{D} D + 2 \bigl(g_s C - e B + 2 g_s^2 |\rho|^2 + m_\rho^2\bigr)\bigr] \rho & = & 0\,.
\label{eq:three}
\eeqn
Equation~\eq{eq:C:2d} along with the conservation law~\eq{eq:drho0},
${\mathrm{Re}} (\bar{\partial} \rho^{(0)}) = 0$, lead to a simple expression for
the transverse component of the field tensor~\eq{eq:f0} of the neutral mesons:
\beqn
C = - i \bar{\partial} \rho^{(0)}\,.
\label{eq:C0}
\eeqn

\subsection{Inhomogeneous condensate of small amplitude}
\label{sec:inhomogeneous}

\subsubsection{Linearized equations of motion}

The classical equations of motions~\eq{eq:cDrho0}--\eq{eq:C0} comprise
a complicated system of equations which is difficult to solve analytically due to the nonlinearities.
However, following our discussion for the GL model (Section~\ref{sec:GL:destructive}),
let us assume that the amplitude of the condensate $\rho$ is very small. Then, the equations
of motion can be linearized and a leading analytical solution can be obtained. The condensate $\rho$ should be small
if the background magnetic field $B_\ext$ exceeds slightly the critical value $B_c$, Eq.~\eq{eq:eBc}. Concretely,
for $B_\ext \geqslant B_c$ we consider the condition,
\beqn
2 g^2_s |\rho|^2 \ll m^2_0\,, \qquad \mbox{or} \qquad \frac{B_\ext}{B_c} - 1 \ll 1\,.
\label{eq:limit}
\eeqn
These relations are analogous to, respectively, weak-condensate conditions \eq{eq:first} and \eq{eq:second} in
the GL model of superconductivity.
We show below that the first and the second relations in \eq{eq:limit} are, in fact, equivalent.

Notice, that Eq.~\eq{eq:cDrho0} coincides with Eq.~\eq{eq:cD:phi} for the order parameter for the ordinary superconductivity
in the GL model~\eq{eq:L:GL} provided that the external magnetic field is close to the critical field~\eq{eq:B:GL:c} of this model.
Therefore we should expect emergence of an analogue of the vortex lattice~\eq{eq:Abrikosov:lattice}
in the $\rho$ system~\eq{eq:L:rho} similarly to the appearance of the Abrikosov lattice~\eq{eq:Abrikosov:lattice} in the GL model.
Thus, the condensate of the $\rho$ mesons in the external magnetic field should definitely be inhomogeneous.
Following the classic example~\cite{ref:Abrikosov}, we consider below the simplest case of the square lattice with the
elementary length~\eq{eq:LB:GL}.

In the weak--condensate regime~\eq{eq:limit} we can work in the leading order in terms of the condensate $\rho_\AS$
(higher order corrections are always omitted below).
Then the equation of motion~\eq{eq:two} gives the following relation:
\beqn
\rho^{(0)}_\AS(x_\perp) = \frac{2 i g_s}{- \partial^2_\perp + m^2_0} \partial |\rho_\AS|^2\,,
\label{eq:rho0:rho}
\eeqn
where
\beqn
\frac{1}{- \partial^2_\perp + m^2_0}(x_\perp) = \frac{1}{2 \pi} K_0(m|x_\perp|)\,,
\eeqn
is the two-dimensional Euclidean propagator of a scalar particle with the mass~$m_0$
and $K_0$ is a modified Bessel function (remember that the subscript ``AS'' stands
for the ``anisotropic superconductor'' solution).

It is very important to notice that Eq.~\eq{eq:rho0:rho} relates the condensate of the neutral $\rho^0$ mesons
with the condensate of the charged $\rho^\pm$ mesons. Thus, if the have an inhomogeneous condensate of the
charged $\rho^\pm$ mesons, then we automatically get the inhomogeneous condensate~\eq{eq:rho0:rho} of the neutral $\rho^0$ mesons as well!
This fact may indicate, that the superconductivity of the $\rho^\pm$ mesons may induce the superfluidity of the $\rho^0$ mesons.
Notice that a relation between the superfluidity of the neutral $\rho^{(0)}$ mesons and the superconductivity of the charged $\rho^\pm$ 
mesons may be guessed from the fact of the vector dominance, Eq.~\eq{eq:rho0:J}.

We interpret
the nonzero condensate~\eq{eq:rho0:rho} as a ``superfluid'' because of the complex nature of the field~$\rho^{(0)}$. Moreover, if,
for a moment, we assume that this field is homogeneous (i.e., coordinate-independent) then rotations of the system around the magnetic field
axis~\eq{eq:rot:invariance:rho} would transform it as a usual complex field in simplest bosonic theories of superfluidity~\cite{ref:Landau},
$\rho^{(0)} \to e^{i \varphi} \rho^{(0)}$. The inhomogeneities of the condensate~\eq{eq:rho0:rho} break spontaneously this global group
down to a discrete group of the rotations of the vortex lattice.

We would like also to notice an important role of the inhomogeneities in the charged $\rho^\pm$ condensate for the superfluidity.
In Section~\ref{sec:homogeneous} we have seen that the homogeneous condensate of the charged $\rho^\pm$ mesons alone is unable to
induce the superfluidity of the $\rho^{(0)}$ mesons: a uniform nonzero expectation value of $\rho^\pm$ does not imply $\rho^{(0)} \neq 0$.
However, the inhomogeneous charged condensate of $\rho^\pm$ automatically induces the inhomogeneous neutral condensate of $\rho^{(0)}$
as one can see from the presence of the derivative $\partial$ in the numerator in the right hand side of Eq.~\eq{eq:rho0:rho}.

The transverse component of the strength tensor~\eq{eq:C:2d} of the neutral $\rho^0$ mesons is
given by Eq.~\eq{eq:C0}:
\beqn
C_\AS(x_\perp) = - i {\bar \partial} \rho^{(0)}_\AS
= 2 g_s \frac{\partial^2_\perp}{- \partial^2_\perp + m^2_0} |\rho_\AS|^2\,.
\label{eq:C:rho}
\eeqn
Due to the identity,
\beqn
\int \dd^2 x_\perp\, \frac{\partial^2_\perp}{- \partial^2_\perp + m^2_0}(x_\perp - y_\perp) = 0\,,
\eeqn
the total ``flux'' of the neutral $\rho^0$ mesons through the transverse plane is always zero,
\beqn
\int_\cA \dd^2 x_\perp \, {(f_{12}^{(0)})}_\AS(x_\perp) \equiv \int_\cA \dd^2 x_\perp \, C_\AS(x_\perp) = 0\,,
\label{eq:C:0}
\eeqn
where the integral is taken over a unit cell $\cA$ of the periodic structure of the ``$\rho$ vortices''.

Next, Eq.~\eq{eq:one} gets simplified,
\beqn
\partial \Bigl(B - \frac{2 e m_0^2}{- \partial_\perp^2 + m^2_0} |\rho_\AS|^2\Bigr) = 0\,,
\eeqn
and its solution becomes as follows
\beqn
B_\AS(x_\perp) = B_\ext + \frac{2 e m_0^2}{- \partial_\perp^2 + m^2_0} |\rho_\AS|^2 - 2 e \, {(\overline{|\rho_\AS|^2})}_\cA \,.
\qquad
\label{eq:B:solution}
\eeqn
Here the last term
\beqn
{(\overline{|\rho_\AS|^2})}_\cA = \frac{1}{L_B^2} \int_\cA \!\dd^2 y_\perp |\rho_\AS(y_\perp)|^2\,,
\label{eq:rho2:mean}
\eeqn
is ``the mean-cell value'' of the condensate squared $|\rho_\AS|^2$.
Due to the identity,
\beqn
\int \dd^2 x_\perp\, \frac{m^2_0}{- \partial^2_\perp + m^2_0}(x_\perp - y_\perp) = 1\,,
\eeqn
the last term in Eq.~\eq{eq:B:solution} guarantees the conservation of the net magnetic flux
through each elementary cell $\cA$:
\beqn
\int_{\cA} \dd^2 x_\perp \, B_\AS(x_\perp) = \int_{\cA} \dd^2 x_\perp \, B^{\mathrm{ext}}
\equiv L_B^2 B^{\mathrm{ext}} = \frac{2 \pi}{e}\,, \qquad
\label{eq:B:conservation}
\eeqn
[here we have used Eq.~\eq{eq:LB:GL}]. The quantization of the magnetic flux~\eq{eq:B:conservation}
is similar to the quantization of the flux of the Abrikosov vortex~\eq{eq:quantized}.

Finally, Eqs.~\eq{eq:three}, \eq{eq:C:0}, \eq{eq:B:solution} and \eq{eq:B:conservation} give us
\beqn
|\rho_{AS}|_{{}_\cA} \equiv {(\overline{|\rho_\AS|^2})}_\cA^{\frac{1}{2}} =
\left\{\!\!
\begin{array}{ccl}
\sqrt{\frac{e(B_\ext - B_c)}{2 g^2_s}}\,, & \ & B_\ext \geqslant B_c,\\
0, & \ & B_\ext < B_c\,,
\end{array}
\right.\quad
\label{eq:norm:rho2}
\eeqn
for the mean value~\eq{eq:rho2:mean} of the condensate. The mean-cell value of the condensate~\eq{eq:norm:rho2}
is shown in Fig.~\ref{fig:condensate}. Notice, that the mean-cell value of the condensate \eq{eq:norm:rho2}
coincides with the value of the uniform condensate~\eq{eq:norm:rho2:h} obtained in the homogeneous-field approximation.

Equation~\eq{eq:norm:rho2} has a few interesting properties. Firstly, this equation represents a typical
behavior of an order parameter. Secondly, Eq.~\eq{eq:norm:rho2} suggests that the
phase transition, which separates the superconducting and the nonsuperconducting phases at $B_\ext = B_c$,
is of a second order (as it is seen clearly in Fig.~\ref{fig:condensate}). And thirdly, Eq.~\eq{eq:norm:rho2} proves the equivalence
between the first and the second conditions of the weak-condensate regime, Eq.~\eq{eq:limit}.

Concluding this section we would like to stress that here we have introduced the new topological object, the ``$\rho$ vortex'',
which is the vortex made of the superconducting $\rho^\pm$ mesons and superfluid $\rho^0$ mesons. This unit vortex-cell
carries the nonzero quantized flux of the magnetic field~\eq{eq:B:conservation} and zero $\rho^0$--flux~\eq{eq:C:0}.
The lattice of such vortices is a ground state of the superconductivity of the QCD vacuum at strong magnetic field.
We discuss this lattice state in details in the next section.

\subsubsection{Inhomogeneous condensate: $\rho$-vortex lattice}

In the regime~\eq{eq:limit} the degree of the inhomogeneity of the magnetic field
$\delta B(x_\perp) = B_\AS(x_\perp) - B_\ext$ in the superconducting state is extremely small.
Indeed, according to Eq.~\eq{eq:B:solution},
\beqn
|\delta B| \sim 2 e |\rho|^2 \ll \frac{e m^2_0}{g_s^2} \approx \frac{e^2}{g_s^2} B_\ext \ll B_\ext\,.
\eeqn
Thus, the inhomogeneity of the magnetic field $\delta B$
is suppressed both by the small amplitude of the condensate~\eq{eq:limit} and by the very small
factor $e^2/g^2_s = 8.8\times 10^{-3}$. From Eqs.~\eq{eq:C:rho} and \eq{eq:B:solution} one also finds that
the stress tensor of the neutral bosons~\eq{eq:C:2d} is small compared to the magnetic field~\eq{eq:B:2d}, $|C| \ll (e/g_s) B_\ext$.

Therefore we can set below $B(x) \simeq B_\ext$ with the very good accuracy.
Then,
\beqn
\cD \simeq \cD_\ext = \partial - e A_\ext = \partial + \frac{e B}{2} z
\eeqn
so that the solution of Eq.~\eq{eq:cDrho0} is
\beqn
\rho_\AS(z) = e^{-\frac{e B}{4} |z|^2} H_\AS(z/L_B)\,,
\label{eq:rho:vac}
\eeqn
where $H_\AS(z)$ is arbitrary analytic function of the argument $z$
and the inter-vortex distance $L_B$ is given in Eq.~\eq{eq:LB:GL}. Following
the known solution~\eq{eq:Abrikosov:lattice} in the conventional
superconductivity~\cite{ref:Abrikosov}, we choose the square form of the lattice cells.
For such periodic structure one gets
\beqn
H_\AS(z) = \sqrt{\frac{e(B_{\mathrm{ext}} - B_c)}{\sqrt{2} g^2_s}} \, e^{- \frac{\pi}{2} z^2}
\sum_{n = -\infty}^{+ \infty} e^{- \pi n^2 + 2 \pi n z}\,, \qquad
\label{eq:H}
\eeqn
where the prefactor was determined with the help of the normalization relation~\eq{eq:norm:rho2} supplemented
by the explicit expressions~\eq{eq:rho:vac} and \eq{eq:H}.

We already know that the homogeneous condensate locks the rotational and gauge degrees of freedom~\eq{eq:locking}.
The inhomogeneities in the condensate break the locked subgroup~\eq{eq:locking} further
down to a discrete subgroup of the lattice rotations $G^{\mathrm{lat}}_{\mathrm{locked}}$:
\beqn
U(1)_{\mathrm{e.m.}} \times O(2)_{\mathrm{rot}} \to U(1)_{\mathrm{locked}} \to G_{\mathrm{locked}}^{\mathrm{lat}}\,.
\label{eq:locking2}
\eeqn
The discrete group $G_{\mathrm{locked}}^{\mathrm{lat}}$ depends on the lattice structure formed by the vortices.

Similarly to the mixed state of the ordinary type--II superconductivity, the $\rho$-vortex centers are located
at the points~\eq{eq:vort:locations}, where the condensate~$\rho_\AS$ vanishes. In the vicinity of the $\rho$-vortex
centers the condensate~\eq{eq:rho:vac} follows the typical Abrikosov-vortex behavior~\eq{eq:Phi:vort}. However,
there are many essential dissimilarities between the vortex systems in the GL model and in the system of the
condensed $\rho$ mesons.

In Fig.~\ref{fig:lattice} we visualize four elementary lattice cells of the $\rho$ vortex lattice in
the transverse plane. We take the external magnetic field with the strength $e B_\ext = (800\,\mbox{MeV})^2 > eB_c$,
so that the system is already in the superconducting state. The strength of the field satisfies the weak-condensate
condition~\eq{eq:limit}. The magnetic length and the elementary distance between the vortices in the square vortex
lattice are, respectively\footnote{Note that the magnetic length $\ell_B$ is of the order of the size of the $\rho$ meson itself,
$r_\rho \sim m_\rho \simeq 0.25\,\mathrm{fm}$. Thus, at these magnetic fields the $\rho$ mesons should mutually overlap
similarly to the overlapping Cooper pairs in the conventional superconductivity. However, regarding the success of the phenomenological
GL model of the superconductivity we don't question the applicability of the phenomenological DSGS model~\eq{eq:L:rho} in the
strong-field regime.}~\eq{eq:LB:GL}, $\ell_B = 0.25\,\mathrm{fm}$ and $L_B = 0.63\,\mathrm{fm}$.
The mean value of the condensate~\eq{eq:norm:rho2} of the $\rho^\pm$ mesons is
$|\phi_\AS| \simeq 23\,{\mathrm{MeV}}$. In Fig.~\ref{fig:lattice} we plot various quantities that characterize the vortex:
the amplitudes of the superconducting and superfluid condensates, the excess of the magnetic field with respect to the
external magnetic field and the field strength of the neutral meson field $C$. One can clearly see that:

\begin{figure}[!thb]
\begin{center}
\vskip 3mm
\includegraphics[scale=0.32,clip=true]{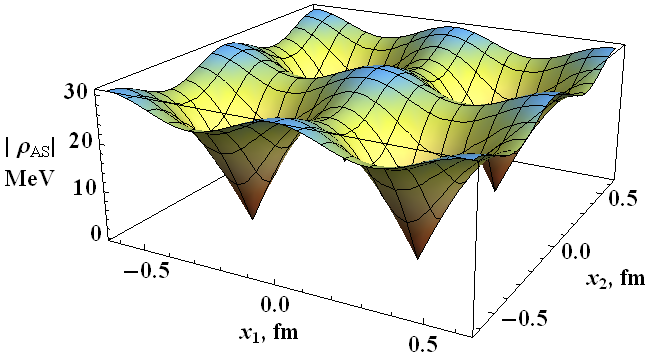}\\
{\large\bf (a)}\\[2mm]
\includegraphics[scale=0.35,clip=true]{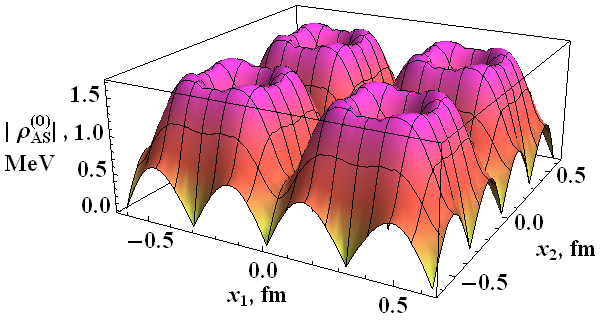}\\
{\large\bf (b)}\\[2mm]
\includegraphics[scale=0.34,clip=true]{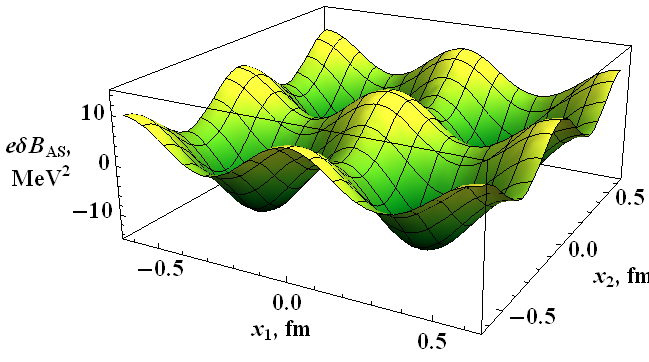}\\
{\large\bf (c)}\\[2mm]
\includegraphics[scale=0.36,clip=true]{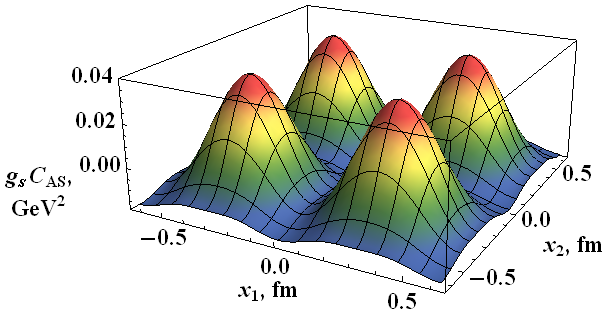}\\[1mm]
{\large\bf (d)}
\end{center}
\vskip -4mm
\caption{Four elementary cells of the $\rho$-vortex lattice in the plane $x_\perp=(x_1,x_2)$. The plane is perpendicular to the 
external magnetic field with the strength $e B_\ext = (800\,\mbox{MeV})^2$. From top to bottom:
(a) the amplitude of the superconducting condensate $\rho$, Eqs.~\eq{eq:rho:vac} and \eq{eq:H};
(b) the amplitude of the superfluid condensate $\rho^{(0)}$, Eq.~\eq{eq:rho0:rho};
(c) the excess of the magnetic field $\delta B(x_\perp) \equiv B(x_\perp) - B_\ext$, Eq.~\eq{eq:B:solution};
(d) the field strength $C$ of the superfluid condensate $\rho^{(0)}$, Eq.~\eq{eq:C:rho}.}
\label{fig:lattice}
\end{figure}

\begin{figure}[!thb]
\begin{center}
\includegraphics[scale=0.4,clip=true]{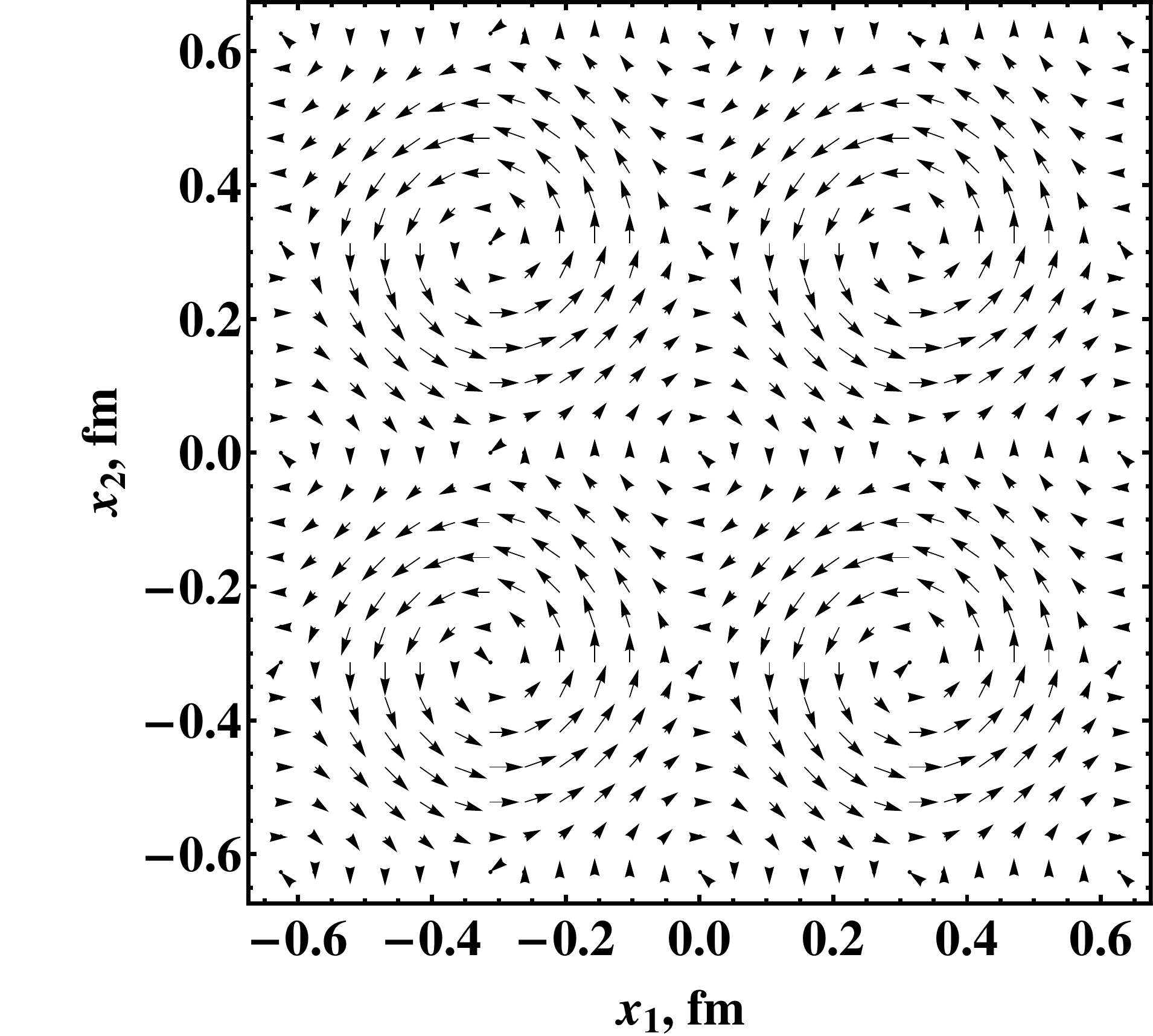}
\end{center}
\vskip -4mm
\caption{The transverse components $J_1$ and $J_2$ of the electric current $J$, Eq.~\eq{eq:Jmu},
in the transverse plane $x_\perp=(x_1,x_2)$. Four elementary cells of the $\rho$-vortex lattice
at the external magnetic field with $e B_\ext = (800\,\mbox{MeV})^2$ are shown.}
\label{fig:j}
\end{figure}

\begin{itemize}

\item The superconducting condensate $\rho$ of the charged vortices $\rho^\pm$, Eqs.~\eq{eq:rho:vac} and \eq{eq:H},
vanishes at the centers of the vortices~\eq{eq:vort:locations}, Fig.~\ref{fig:lattice}(a). In the vortex core the amplitude of
the condensate $|\rho|$ is a linear function of the distance from the vortex center. This feature is
similar to the behavior of the condensate near a typical Abrikosov vortex with a unit vorticity~\eq{eq:Phi:vort}.

\item The superfluid condensate $\rho^{(0)}$, Eq.~\eq{eq:rho0:rho},
has a toothlike structure, Fig.~\ref{fig:lattice}(b). It vanishes at the locations of all
local extrema of the superconducting condensate including the centers of the vortices. The amplitude of the superfluid condensate
is maximal at the points of steepest behavior of the superconducting condensate, Fig.~\ref{fig:lattice}(a).

\item The magnetic field strength $B$, Eq.~\eq{eq:B:solution},
takes its minimal values at the centers of the vortices, Fig.~\ref{fig:lattice}(c). The maxima of $B$ are located outside the vortex cores.
This feature contradicts our intuition: in the ordinary superconductivity the strength of the magnetic field is maximal at
the center of the Abrikosov vortex. In fact, the $\rho^\pm$ condensate has its own magnetic dipole moment due to the large, $g=2$, gyromagnetic
ratio of the $\rho$ vortex. This dipole moment contributes only to the magnetic field outside the vortex cores, where the condensate
of the $\rho^\pm$ condensate is large, Fig.~\ref{fig:lattice}(a). The electric current $J$, Eq.~\eq{eq:Jmu}, is visualized in Fig.~\ref{fig:j}.

\item The strength of the neutral meson field $C$ of the superfluid [Eq.~\eq{eq:C:rho}] takes its maxima at the locations of
the $\rho$ vortices, Fig.~\ref{fig:lattice}(d). Thus, the
$\rho$ vortices share this important property of the ordinary superfluid vortices as well.

\end{itemize}

Summarizing, the vortex core expels both superconducting and superfluid condensates of the charged and neutral $\rho$ mesons, respectively.
The magnetic field takes its maxima outside the vortices, while the strength of the superfluid (electrically neutral) field
is peaked at the vortex centers.

\subsubsection{Anisotropic superconductivity}

The basic property of a superconductor is the absence of the resistivity. This feature is reflected, in particular,
in the first London equation~\eq{eq:London:GL} in the GL model.

There is a simple way to derive analogues
of the London equations for the condensed state of $\rho$ mesons
in the external magnetic field. First, we notice that Eqs.~\eq{eq:rho0:2} and \eq{eq:rho0:J} imply:
\beqn
\bigl(\partial^\alpha \partial_\alpha + m^2_0 \bigr) \partial_{[\mu} J_{\nu]} = m_0^2 \partial_{[\mu} J^{\mathrm{ch}}_{\nu]}\,.
\label{eq:ddJ}
\eeqn
Then, we take $\mu=0$ and $\nu=3$ in Eq.~\eq{eq:ddJ} and use Eq.~\eq{eq:Jch} to get expressions for the $\mu=0,3$ components of
the charged currents:
\beqn
J^{\mathrm{ch}}_a & = & 2 i e \bigl[\rho^* D_a \rho - (D_a \rho)^* \rho \bigr]\,, \quad a=0,3\,.
\label{eq:Jch:a}
\eeqn
Following the logic of the derivation of the London equation~\eq{eq:London:GL}
in the Ginzburg--Landau approach (Section~\ref{sec:GL:homogeneous}), one gets from Eqs.~\eq{eq:ddJ} and \eq{eq:Jch:a}:
\beqn
\frac{\partial J_3 (x_0, x_\perp)}{\partial x_0} = - 4 e^2 h^2_\AS(x_\perp) E_3\,,
\label{eq:London:3}
\eeqn
where $x_\perp = (x_1,x_2)$. The inhomogeneous quantity
\beqn
h^2_\AS = \frac{m_0^2}{- \partial^2_\perp + m_0^2} |\rho_\AS|^2\,.
\label{eq:h2}
\eeqn
plays the r\^ole of the $|\Phi_0|^2$ condensate or $m^2_A/e^2$ in the conventional London relation~\eq{eq:London:GL}.

Equation~\eq{eq:London:3} implies that the $\rho$-meson condensate exhibits the superconductivity phenomenon along
the direction of the external magnetic field $\vec B$: the electric current growing linearly with time if a weak external
electric field is applied.

Notice, that due to the periodicity of the inhomogeneous condensed state
the mean-cell values of the squares of the ``effective condensate''~\eq{eq:h2} and
of the real condensate \eq{eq:rho2:mean} coincide identically:
\beqn
{(\overline{h^2_\AS})}_\cA \equiv {(\overline{|\rho_\AS|^2})}_\cA\,.
\label{eq:rho2:h2}
\eeqn
Averaging Eq.~\eq{eq:London:3} over an elementary square cell in transverse directions and using Eq.~\eq{eq:norm:rho2}
we get the cell-averaged value of the electric current $(\overline{J}_{3})_{\cA}$:
\beqn
\frac{\partial}{\partial t}
(\overline{J}_{3})_{\cA}
= - \frac{2 e^3}{g^2_s} (B_\ext - B_c) E_3\,, \qquad
\label{eq:London:4}
\eeqn
where $B_\ext > B_c$ and we assumed, as usual, that the external electric field~$E_3$ is a space-time independent quantity. The
longitudinal (i.e., directed along $\vec B_\ext$)
superconductivity sets in as the external field $B_\ext$ exceeds the critical value $B_c$, Eq.~\eq{eq:eBc}.

It is easy to prove that the superconductivity phenomenon has an anisotropic nature: in the transverse
(i.e., perpendicular to $\vec B_\ext$) directions
the superconductivity is absent. In order to prove this fact let us apply a weak spacetime-independent
electric field $\vec E_\ext = (E_{\ext,1}, E_{\ext,2}, 0)$ perpendicularly to the strong magnetic field background
${\vec B}_\ext  = (0,0,B_\ext)$.
This electric field should test a possible transverse superconductivity of the $\rho^\pm$-meson condensate
which could also be created by the strong magnetic field.

In order to show that the $\vec B_\ext$-transverse electric field does
not create an accelerating electric current, we notice that an appropriate Lorentz boost may transform this system
of the nonparallel $\vec E_\ext$ and $\vec B_\ext$ fields into the frame where the electric field is zero, $\vec E'_\ext = 0$. Obviously, in the
new frame there are no linearly growing electric currents, so that in the initial frame such runaway currents are absent as well and
\beqn
\frac{\partial J_i (x_0, x_\perp)}{\partial t} = 0\,, \qquad
\frac{\partial}{\partial t} (\overline{J}_i)_{\cA} = 0\,,
\qquad i = 1,2\,. \qquad
\label{eq:London:1:2}
\eeqn
This argument does not work for the parallel electric and magnetic fields which were used to prove the
longitudinal superconductivity~\eq{eq:London:3}. Indeed, in this case the
scalar product $(\vec E_\ext\cdot \vec B_\ext) \propto \varepsilon_{\mu\nu\alpha\beta} F^{\mu\nu}_\ext F^{\alpha\beta}_\ext$
is a Lorentz-invariant quantity which is insensitive to boosts and rotations. Thus, if $\vec E_\ext\parallel \vec B_\ext$
then there is no frame where the external electric field $\vec E'_\ext$ is zero.

Equations~\eq{eq:London:4} and \eq{eq:London:1:2} imply that the (cell-averaged) electric conductivity~\eq{eq:J:London}
contains an anisotropic complex--valued contribution~\eq{eq:sigma:London:1} which has a singular part at $\omega=0$:
\beqn
\sigma^{\mathrm{sing}}_{kl}(\omega) = \frac{\pi e^3}{g^2_s}(B_\ext - B_c) \Bigl[\delta(\omega)+\frac{2i}{\pi \omega}\Bigr]
\delta_{k 3} \delta_{l 3}\,, \quad
\label{eq:sigma:sing}
\eeqn
where the index $i=3$ corresponds to the direction of the external magnetic field $\vec B_\ext$.

The anisotropy of the superconductivity is quite similar to the anisotropy of the ``usual'' conductivity of the quenched QCD
vacuum which was found in lattice simulation in Ref.~\cite{Buividovich:2010tn} for weaker magnetic fields.
An explanation of the anisotropy could be as follows: in a background of a uniform magnetic field the electric charges may
move along the axis of the magnetic field while the motion in the transverse direction is limited to the
spatial size $\ell_B$ of the low Landau orbits~\eq{eq:LB:GL}. In a sufficiently strong magnetic field,
and in absence of scattering of the charge carriers, the net transverse motion of the charges is 
suppressed contrary to the motion in the longitudinal direction.

\subsubsection{Absence of longitudinal Meissner effect}

We have a very unusual situation: in our paper we suggest that in the QCD vacuum the strong magnetic field induces
the superconductivity of $\rho$ mesons, while all our experience in the condensed matter systems tells
us that we should expect the opposite phenomenon~\cite{ref:Landau,ref:Abrikosov}: the external magnetic field should
destroy the superconductivity due to the Meissner effect (Section~\ref{sec:GL:Meissner}). In order to find a reason
for this would-be inconsistency between the usual superconductor and the $\rho$-meson system
let us apply the considerations of Section~\ref{sec:GL:Meissner} to the $\rho$ mesons.

According to the Maxwell equations the electric currents that could screen the
external magnetic field ${\vec B}_{\ext} = (0,0,B_\ext)$ should circulate in the transverse $x_\perp$-plane.
In turns, the superconducting current in the transverse plane, $J_\AS \equiv J_{\AS,1} + i J_{\AS,2}$, can be related to the
neutral meson current~\eq{eq:rho0:rho} via the vector dominance relation~\eq{eq:rho0:J}:
\beqn
J_\AS(x_\perp) = \frac{e m^2_0}{g_s} \rho^{(0)}_\AS(x_\perp)
= \frac{2 i e m^2_0}{- \partial^2_\perp + m^2_0} \partial |\rho_\AS|^2\,.
\eeqn
Then in the system of the condensed $\rho$ mesons, the analogue of the second London equation~\eq{eq:London:GL:2}
for the longitudinal magnetic field
can be written as follows (here we use the relation ${\bar \partial} \partial = \partial^2_\perp$):
\beqn
({\vec \partial} \times {\vec J}_\AS)_3 \equiv {\mathrm{Im}} ({\bar \partial} J_\AS)
= 2 e m^2_0 \frac{\partial^2_\perp}{- \partial^2_\perp + m^2_0} |\rho_\AS|^2\,. \quad
\label{eq:London:3:rho}
\eeqn
The right hand side of this equation depends on the external magnetic field $B_\ext$
via the superconducting density $\rho_\AS$, Eq.~\eq{eq:rho:vac}.

Equations~\eq{eq:London:3:rho}, \eq{eq:rho:vac} and \eq{eq:H} provide us with an implicit
expression for the curl of the screening currents. However, even without knowledge of the
explicit form of these solutions one can show that these transverse currents both {\it screen and enhance}
the external magnetic field in such a way that the net effect in one elementary vortex cell is
precisely zero. Indeed, let us integrate left and right hand sides of Eq.~\eq{eq:London:3:rho}
over an elementary unit cell, take into account the periodicity of the solution~\eq{eq:rho:vac}
and use the following property
\beqn
\int \dd^2 x_\perp\, \frac{\partial^2_\perp}{- \partial^2_\perp + m^2_0}(x_\perp - y_\perp) = 0\,.
\eeqn
Thus, the cell-averaged right hand side of the second London equation~\eq{eq:London:3:rho}
for $\rho$ mesons is zero,
\beqn
\int\limits_\cA \dd^2 x_\perp \, ({\vec \partial} \times {\vec J}_\AS)_3  = 0\,, \quad \mbox{[condensed $\rho^\pm$ mesons]}\,.
\qquad
\label{eq:London:rho:int}
\eeqn
while in the GL model the same procedure would give us the constant quantity
in the right hand side of \eq{eq:London:GL:2}:
\beqn
\int\limits_\cA \dd^2 x_\perp \, ({\vec \partial} \times {\vec J}_{\mathrm{GL}})_3  = - m^2_A B_\ext\,,
\quad \mbox{[GL model]}\,. \qquad
\label{eq:London:GL:int}
\eeqn
This fact simply means that in the state of the condensed $\rho$ mesons, the external magnetic
field induce the transverse superconducting currents which are circulating both clockwise and
counterclockwise contrary. Consequently, the external magnetic field is enhanced in some regions
of the transverse plane and it is suppressed in the other regions. Contrary to the ordinary
superconductor, the net current circulation of the superconducting $\rho$ currents per a unit
lattice cell is exactly zero~\eq{eq:London:rho:int}, while in the ordinary superconductor the
net circulation is a linearly growing function of the external magnetic field.

Thus, we have found that the external magnetic field of any strength $B_\ext > B_c$ does not
experience the screening inside the $\rho$-superconductor: the magnetic flux propagates
freely inside the superconductor. The same statement is not true for
the ordinary superconductor in the purely superconducting state: the magnetic field tries
to avoid the superconductor (the Meissner effect). Thus, in a loose sense one can interpret the absence of the
net circulating currents~\eq{eq:London:rho:int} as the absence of the ``longitudinal'' Meissner effect.

On the other hand, our system is very similar to the ordinary Abrikosov lattice in the mixed state of
the type-II superconductor,  Section~\ref{sec:GL:Abrikosov}:
in the mixed state the magnetic field forms an inhomogeneous state and propagate though the superconductor,
basically, in the cores of the Abrikosov vortices. In this case, however, the external magnetic field must be bounded
both from above and from below, contrary to our $\rho$--superconductivity in the QCD vacuum.

One can try to address the question about the existence of the Meissner effect in the $\rho$-superconductor in a different way.
In the ordinary superconductivity the Meissner effect is usually formulated as follows: if we apply
a weak ``test'' magnetic field, say ${\vec B}'_\ext = (B'_\ext,0,0)$,
along the boundary of a superconductor then this field will be
screened inside the superconductor according to Eq.~\eq{eq:Meissner}, i.e. ${\vec B}(x_3) = (e^{-m_A x_3} B'_\ext,0,0)$.
This experiment, however, is senseless in the case of the $\rho$--condensation
because this condensation is induced in the rotationally-invariant vacuum by the magnetic field itself.
Indeed, assume that we have a combination of the two external magnetic fields: the strong field $\vec B'_\ext$, which
induces the conductivity, and the additional weak field $\vec B''_\ext$ which is superimposed onto $\vec B'_\ext$ transversely
$(\vec B'_\ext \cdot \vec B''_\ext) = 0$, in order to check the Meissner effect. Due to vacuum environment it is clear
that the sole r\^ole of the additional field $\vec B''_\ext$ is to rotate the primary field $\vec B'_\ext$.
After simple rotation of our coordinate around its origin we get a new field $\vec B_\ext = \vec B'_\ext + \vec B''_\ext$
so that the role of the additional ``test'' field is to rotate the directions of the $\rho$-vortices in the condensed state.
Thus, the question of the (non)existence of the transverse Meissner effect cannot be formulated in a selfconsistent way.

\section{Conclusions}

\begin{figure}[!thb]
\begin{center}
\includegraphics[scale=0.47,clip=true]{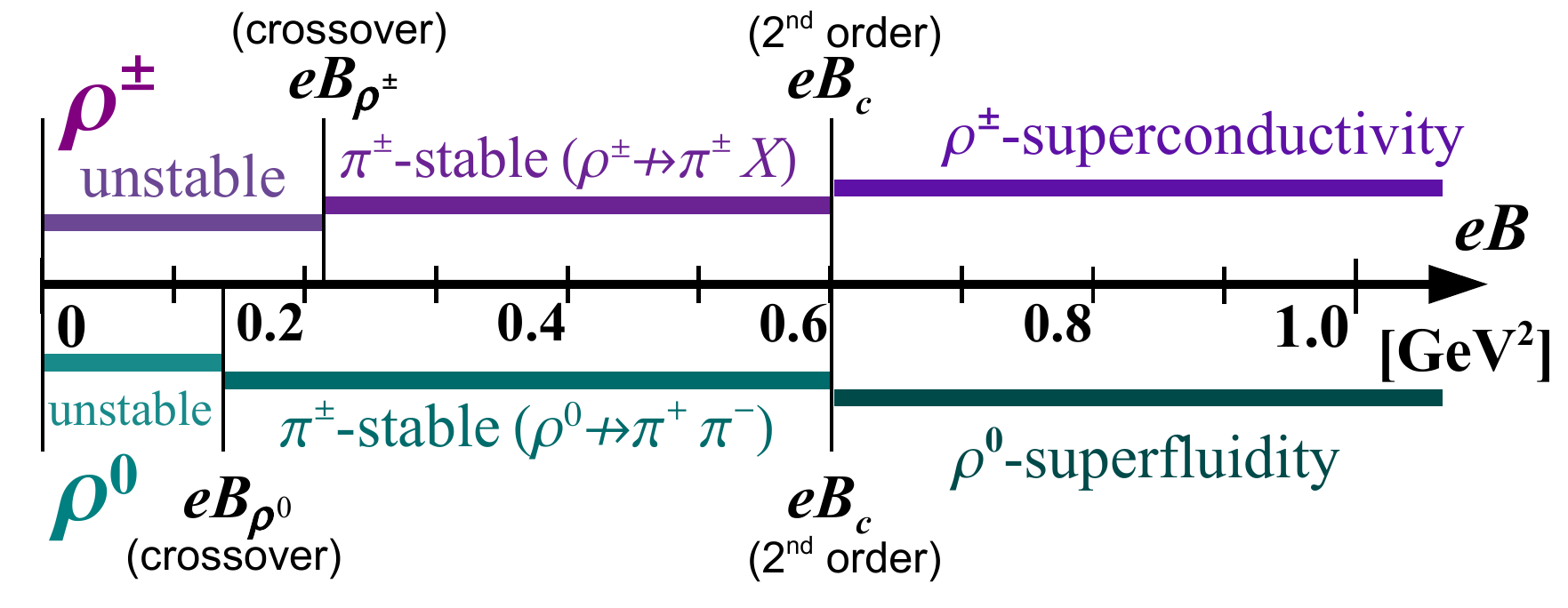}
\end{center}
\vskip -5mm
\caption{The expected phase diagram: the impact of strong
external magnetic field $B \equiv B_{\ext}$
on the $\rho$-meson degrees of freedom in the QCD vacuum.}
\label{fig:diagram}
\end{figure}

We argue that in a sufficiently strong background magnetic field the QCD vacuum
may undergo a spontaneous transition to a superconducting state via condensation
of the charged $\rho^\pm$ mesons. The critical strength of the magnetic field is given in Eq.~\eq{eq:eBc}.
The superconductivity is understood in the
usual electromagnetic sense. Moreover, unlike the color superconductivity, the superconducting QCD state is
suggested to be formed in the cold vacuum, i.e. at zero temperature and at zero chemical potentials.
Our vision of the phase diagram of the cold QCD vacuum in terms of the $\rho$-meson
degrees of freedom is illustrated is Fig.~\ref{fig:diagram}.

We have found the following basic properties of the superconducting state:

\begin{enumerate}

\item The superconducting effect occurs because of the nonminimal
coupling of the charged $\rho$ mesons to the electromagnetic field. The strong magnetic field
enhances the electromagnetic superconductivity of the QCD vacuum instead of destroying it.

\item Due to simple kinematical reasons a strong enough magnetic field makes the
lifetime of the $\rho$ mesons much longer by closing the dominant decay channels
($\rho^\pm \to \pi^\pm \pi^0$ and $\rho^0 \to \pi^+ \pi^-$) of the $\rho$ mesons into the charged pions.
The estimations of the corresponding critical field strengths for the charged and neutral $\rho$ mesons
are given in Eqs.~\eq{eq:B:rho} and \eq{eq:B:rho:0}. Since these critical strengths are smaller than the
critical superconducting field~\eq{eq:eBc}, the superconducting condensate should be intrinsically stable, at least at
the scale of the strong interactions.

\item The transitions between the unstable and stable regions
of the $\rho$ mesons are expected to be smooth crossovers
while the onset of the superconductivity
is expected to be a second order phase transition.

\item The superconducting state is anisotropic: the electric resistance is zero
only along the axis of the magnetic field.

\item The superconducting state is inhomogeneous: the condensate shares similarity
with the Abrikosov vortex lattice in the mixed state of a type-II superconductor.

\item The pure homogeneous superconducting state is not formed.

\item The onset of the superconductivity of the charged $\rho^\pm$ mesons leads to emergence of an
inhomogeneous superfluidity of the neutral $\rho^0$ mesons. The superfluidity is induced by the
inhomogeneities of the superconducting condensate.

\item The inhomogeneous superconducting state is realized as the $\rho$-vortex lattice.
Locally, the $\rho$-vortex core expels both superconducting and superfluid condensates of the charged and neutral $\rho$ mesons, respectively.
The magnetic field takes its maxima outside the vortices, while the
the strength of the superfluid (electrically neutral) field is peaked at the vortex centers.
However, the unit $\rho$--vortex cell carries one unit of the quantized magnetic flux of the magnetic field while the net $\rho^0$ ``flux'' is vanishing.

\item The spontaneous emergence of the superconducting condensate ``locks'' the rotations of the system around the magnetic
field axis with a global subgroup of the gauge transformations. The inhomogeneities of the condensate break the locked
group further to the group of discrete rotations of the vortex lattice.

\item The Meissner effect (understood in the usual sense) cannot be realized in the superconducting QCD state due to
the Lorenz-invariance of the vacuum.

\end{enumerate}

Our results also imply that the inhomogeneous Ambj\o rn--Olesen state~\cite{Ambjorn:1988gb,Ambjorn:1988tm} of the vacuum of the
electroweak model is, in fact, an anisotropically superconducting state. This state may in principle be realized in first moments of the Universe
if strong enough magnetic fields are created in the primordial era~\cite{Grasso:2000wj}.
The superconducting nature of the Ambj\o rn--Olesen state may have imprints in the
large-scale structure of the magnetic fields in the present-day Universe.

\acknowledgments

This work has been partially supported by the French Agence Nationale de la Recherche project ANR-09-JCJC ``HYPERMAG''.

\end{document}